\documentclass[11pt]{article}

\usepackage{amscd,amstex,amssymb}

\renewcommand{\Bbb}{\mathbb}
\newcommand{\goth}{\mathfrak}
\newcommand{\arrow}{{\:\longrightarrow\:}}
\newcommand{\Z}{{\Bbb Z}}
\newcommand{\C}{{\Bbb C}}
\newcommand{\R}{{\Bbb R}}

\newcommand{\6}{\partial}
\newcommand{\1}{\sqrt{-1}\:}
\newcommand{\g}{{\goth g}}

\newcommand{\restrict}[1]{{\left|_{{\phantom{|}\!\!}_{#1}}\right.}}

\renewcommand{\c}[1]{{\cal #1}}
\newcommand{\calo}{{\cal O}}

\renewcommand{\phi}{\varphi}
\renewcommand{\epsilon}{\varepsilon}
\renewcommand{\geq}{\geqslant}
\renewcommand{\leq}{\leqslant}
\newcommand{\even}{{\rm even}}
\newcommand{\odd}{{\rm odd}}
\newcommand{\im}{\operatorname{im}}
\newcommand{\Aff}{\operatorname{Aff}}

\newcommand{\comment}[1]{{}}

\def\blacksquare{\hbox{\vrule width 4pt height 4pt depth 0pt}}

\makeatletter

\newcommand{\ps@verbit}{%
  \renewcommand{\@oddhead}{%
          \scriptsize
          {Mirror conjecture}
          \hfil\tiny {final version, November 95}}
  \renewcommand{\@evenhead}{\@oddhead}
  \renewcommand{\@oddfoot}{\hfil\thepage\hfil}
  \renewcommand{\@evenfoot}{\@oddfoot}}
 
\newcounter{Mycounter}[section]
\newcounter{lemma}[section]
\renewcommand{\thelemma}{{Lemma \thesection.\arabic{lemma}}}
\newcommand{\lemma}{%
     \setcounter{lemma}{\value{Mycounter}}
     \refstepcounter{lemma}
     \stepcounter{Mycounter}
     {\bf \thelemma:\ }}

\newcounter{claim}[section]
\renewcommand{\theclaim}{{Claim \thesection.\arabic{claim}}}
\newcommand{\claim}{%
     \setcounter{claim}{\value{Mycounter}}
     \refstepcounter{claim}
     \stepcounter{Mycounter}
     {\bf \theclaim:\ }}

\newcounter{sublemma}[section]
\renewcommand{\thesublemma}{{Sublemma \thesection.\arabic{sublemma}}}
\newcommand{\sublemma}{%
     \setcounter{sublemma}{\value{Mycounter}}
     \refstepcounter{sublemma}
     \stepcounter{Mycounter}
     {\bf \thesublemma:\ }}

\newcounter{corollary}[section]
\renewcommand{\thecorollary}{{Corollary \thesection.\arabic{corollary}}}
\newcommand{\corollary}{%
     \setcounter{corollary}{\value{Mycounter}}
     \refstepcounter{corollary}
     \stepcounter{Mycounter}
     {\bf \thecorollary:\ }}

\newcounter{theorem}[section]
\renewcommand{\thetheorem}{{Theorem \thesection.\arabic{theorem}}}
\newcommand{\theorem}{%
     \setcounter{theorem}{\value{Mycounter}}
     \refstepcounter{theorem}
     \stepcounter{Mycounter}
     {\bf \thetheorem:\ }}

\newcounter{proposition}[section]
\renewcommand{\theproposition}
       {{Proposition \thesection.\arabic{proposition}}}
\newcommand{\proposition}{%
     \setcounter{proposition}{\value{Mycounter}}
     \refstepcounter{proposition}
     \stepcounter{Mycounter}
     {\bf \theproposition:\ }}

\newcounter{definition}[section]
\renewcommand{\thedefinition}
       {{Definition \thesection.\arabic{definition}}}
\newcommand{\definition}{%
     \setcounter{definition}{\value{Mycounter}}
     \refstepcounter{definition}
     \stepcounter{Mycounter}
     {\bf \thedefinition:\ }}

\newcounter{example}[section]
\renewcommand{\theexample}{{Example \thesection.\arabic{example}}}
\newcommand{\example}{%
     \setcounter{example}{\value{Mycounter}}
     \refstepcounter{example}
     \stepcounter{Mycounter}
     {\bf \theexample:\ }}

\newcounter{remark}[section]
\renewcommand{\theremark}{{Remark \thesection.\arabic{remark}}}
\newcommand{\remark}{%
     \setcounter{remark}{\value{Mycounter}}
     \refstepcounter{remark}
     \stepcounter{Mycounter}
     {\bf \theremark:\ }}

\newcounter{problem}[section]
\newcounter{question}[section]
 \@ifundefined{boxed}{%
     }{}

\@addtoreset{equation}{section}
\@addtoreset{footnote}{section}
\makeatother

\begin{document}

\begin{center}
{\Large\bf
	Mirror Symmetry for hyperk\"ahler manifolds.}\\[4mm]
Misha Verbitsky,\\[4mm]
{\tt verbit@@math.harvard.edu}
\end{center}

\hfill

{\small 
\hspace{0.2\linewidth}\begin{minipage}[t]{0.7\linewidth}
We prove the Mirror Conjecture for 
Calabi-Yau manifolds equipped with a holomorphic symplectic form,
also known as complex manifolds of hyperk\"ahler type.
We obtain that a complex manifold of hyperk\"ahler type is mirror dual to 
itself. The Mirror Conjecture is stated (following Kontsevich,
ICM talk) as the equivalence of certain
algebraic structures related to variations of Hodge structures. 
We compute the canonical flat coordinates on the moduli space
of Calabi-Yau manifolds of hyperk\"ahler type, 
introduced to Mirror Symmetry by Bershadsky, Cecotti, Ooguri and Vafa.
\end{minipage}
}

\hfill


\section{Introduction.}
\label{_Intro_Section_}


By a holomorphically symplectic manifold we understand a 
complex manifold equipped with a closed holomorphic 2-form, which
is non-\-de\-ge\-ne\-ra\-te. A hyperk\"ahler manifold is a Riemannian
manifold equipped with a quaternionic action which is parallel
with respect to a Levi-Civita connection. Every hyperk\"ahler
manifold is complex (the complex structure is induced by any
embedding $\C \hookrightarrow {\Bbb H}$) and holomorphically
symplectic. The converse is also
true in the compact case, as implied by the Calabi-Yau theorem.
For details and basic results on hyperk\"ahler
manifolds, see \cite{_Besse_}, \cite{_main_}.

\subsection{A summary for those who like physics.}

Mirror Symmetry has a rich history, which this article
mostly ignores. For up-to-date references to the physical literature, 
the reader is advised to look in \cite{_Greene_}.

Let $M$ be a compact holomorphically symplectic manifold.
For a generic complex structure on $M$, $M$ admits no holomorphic
curves (\cite{Verbitsky:Symplectic_II_}). This allows one to compute
the (tautological, because no instanton corrections are required)
correlation functions of the A-model, for $M$ as a target space.

The moduli space $Comp$ of $M$ is equipped with a locally biholomorphic map
to a quadric hypersurface $C$ in a complex projective space 
${\Bbb P}H^2(M,\C)$. The space $C$ is equipped with a transitive action
of a group $G_0(M) \cong SO(3, n-3)$, $n = h^2(M)$. 
Consider the variation $V$ of Hodge structures over $Comp$ 
associated with the cohomology of $M$.
We prove that there exists a $G_0(M)$-equivariant
variation of Hodge structures $\underline V$ on $C$ such that 
$V$ is a pullback of $\underline V$. This allows one to compute
$V$ explicitly. Using $V$, we 
compute the correlation functions for the B-model, in terms
of the $G_0(M)$-action.

Usually 
in Mirror Symmetry,
the parameter space for the A-model is  
the complexified K\"ahler cone, but in the case
of hyperk\"ahler manifolds, the correlation functions can be analytically
continued to the whole space 
\linebreak 				
$H^{1,1}(M)$, which contains 
the K\"ahler cone as an open subset.  The parameter space for
the B-model is the moduli space $Comp$, equipped with the canonical
flat coordinates of \cite{_BerCecOogVaf_}. We identify these 
parameter spaces locally using the flat coordinates, and compare
correlation functions, computed explicitly.

\subsection{A summary for those who like mathematics.}

Let $M$ be a compact Calabi-Yau manifold.\footnote{In this article,
by Calabi-Yau manifold we understand a Ricci-flat K\"ahler manifold.}
Physicists associate with $M$ two associative, graded commutative
algebras with unit: the Yukawa algebra, determined by the
complex structure, and the quantum cohomology algebra, which depends
on the K\"ahler class. By definition, the Yukawa algebra of $M$
is isomorphic to the ring $\oplus H^i(\Lambda^j TM)$, where 
$\Lambda^j TM$ is the $j$-th exterior power of the bundle of holomorphic
vector fields, and multiplication is defined naturally by the K\"unneth
formula (Section \ref{_Tia_Todoro_coordi_Section_}). 
Using the triviality of the canonical class of $M$, we obtain
an isomorphism of holomorphic vector bundles 

\[ 
   \eta:\; \Lambda^i TM \arrow \Omega^{n-i}M,
\] 
where $n = \dim_\C M$.
Thus, the Yukawa product can be considered as a multiplicative
structure on the cohomology of $M$:
\footnote{We denote by $\Omega^i(M)$ the 
sheaf of holomorphic $i$-forms on $M$.
Thus, $H^{i}(\Omega^j(M)) = H^{j,i}(M)$.}
\[ \bullet_Y:\; H^{i,j}(M) \times H^{i',j'}(M) \arrow  
   H^{i+i'-n,j+j'}(M) 
\]

The quantum cohomology algebra (\ref{_quantu-coho-Fro_Definition_})
is a deformation of
the usual cohomology algebra defined via counting the
rational curves on $M$. As usual, 
an associative algebra is called {\bf Frobenius} if it is equipped with
a non-degenerate invariant scalar product
(\ref{_Fro_alge_Definition_}). The Yukawa algebra and the Quantum
cohomology algebra are Frobenius, which follows from the definitions.

 The main ingredient
of the Mirror Conjecture is the existence of the so-called ``Mirror dual''
Calabi-Yau manifold $W$.
Mirror Symmetry is often stated as an isomorphism
between Frobenius algebras: the quantum cohomology 
Frobenius algebra associated with $M$ is conjecturally isomorphic to the
Yukawa cohomology  Frobenius algebra of $W$, and vice
versa. In this form, the Mirror Conjecture trivially holds for
$M$ and $W$ being the same holomorphically symplectic
manifold (\ref{_mirror_for_hype_Theorem_}).

The aim of this article is to prove the refined form of the Mirror
Conjecture, which states an isomorphism between certain 
algebraic structures, which we call
``variations of Frobenius algebras.''

A variation of Frobenius algebras (see \ref{_VFA_Definition_} for
the {\it exact} definition) over a base $X$ is a 
variation of Hodge structures $B$ equipped with a structure
of Frobenius algebra on its associated graded bundle $B^{gr}$, 
and a sheaf homomorphism $TX \stackrel \tau \hookrightarrow B^{gr}$,
such that the Kodaira-Spencer map $B^{gr}\otimes TX \arrow B^{gr}$
coinsides with 
$x\otimes \overrightarrow v \arrow x \cdot \tau(\overrightarrow v)$.
Every Calabi-Yau manifold $M$ produces two
variations of Frobenius algebras: the VFA of Yukawa, $Y(M)$
and the Quantum Cohomology\footnote{Quantum VFA is defined only up
to certain convergence assumptions, which are part of the Mirror Conjecture.
These assumptions are satisfied for a compact holomorphically symplectic
manifold which is generic in its deformation class.}
 VFA, $Q(M)$.
In this setting, Calabi-Yau manifolds $M$ and $W$ are 
{\bf Mirror dual} if the VFA $Q(M)$ is locally isomorphic
to $Y(W)$ and $Y(M)$ is locally isomorphic to $Q(W)$,
and these isomorphisms are compatible with flat coordinates
on the moduli space of complex structures, introduced in
\cite{_BerCecOogVaf_}.
We give the precise statement
of the Mirror Conjecture in Section \ref{_Qua_VFA_Section_}.
In a similar form, the Mirror Conjecture is stated by M. Kontsevich
(\cite{_Kontsevich_Zurich_}).

The Mirror Conjecture appears to be true for a wide 
variety of Calabi-Yau manifolds, but not for all Calabi-Yau. 
There is not a single case in which the Mirror Conjecture is proven for
 Calabi-Yau manifolds in the strict 
sense.\footnote{Calabi-Yau manifolds of dimension 
greater than 2 with holomorphic Euler characteristic $\chi(\calo(M))=2$.
In dimension 3, such manifolds were the original source and the testbed
of the Mirror Conjecture.}
There are several approaches to the Mirror Conjecture 
for K3 surfaces and compact tori, which are, by convention,
Mirror self-dual: \cite{_Todorov:K3_}, \cite{_Bogomolov:K3_}, 
\cite{_AM:K3_} (for additional references see \cite{_AM:K3_}).

\subsection{On the equivariance of a variation of Hodge structures.}

Let $M$ be a holomorphic symplectic manifold of K\"ahler type,
with \[ h^{2,0}(M) =1, \ \ h^{1}(M)=0,\] and $Comp$ be
its (coarse, marked) moduli space. We have the {\bf period map}
$P_c:\; Comp \arrow {\Bbb P}H^2(M, \C)$ associating a line
$H^{2,0}_I(M) \subset H^2(M, \C)$ to a complex structure $I$.
There is a canonical non-degenerate symmetric pairing

\[ (\cdot,\cdot)_{\c H}:\; 
   H^2(M, \R)\times H^2(M, \R) \arrow \R 
\]
defined by Beauville (see \cite{_main_}, \cite{_coho_announce_}).
Complexifying $H^2(M, \R)$, we can consider 
$(\cdot,\cdot)_{\c H}$ as a complex-linear, 
complex-valued form on $H^2(M, \C)$. For all $I\in Comp$, the point
$P_c(I)$ belongs to the quadric cone $C \subset {\Bbb P}H^2(M, \C)$,

\[ C = \{ l \;\; |\;\; (l,l)_{\c H} =0\}. \]
The Torelli principle (proved by Bogomolov, \cite{_Bogomolov:78_}) 
implies that the map \[ P_c:\; Comp \arrow C\] is etale.

Let $\underline{\c H}= \oplus H^{p,q}(M)$ be the  
variation of Hodge structures
(VHS) on $Comp$ associated with the total cohomology space of $M$. 
The Results of \cite{_main_} imply that there exists a variation of 
Hodge structures $\c H$ on $C$, such that $\underline {\c H}$
is the pullback of a variation of Hodge structures $\c H$: 
$\underline {\c H} = P^*_c(\c H)$
(also, this is an immediate implication of 
\ref{_eta_depends_on_peri_Proposition_}). The set $C$ is equipped with
a natural action of the group $G_0(M) = SO\bigg(H^2(M, \R), \
(\cdot,\cdot)_{\c H}\bigg)$. 
The group $G_0(M)$ also acts in the total cohomology space $H^*(M)$ 
of $M$ (\cite{_main_}). The main idea used in the proof
of Mirror Symmetry is the following theorem, implicit
in \cite{_main_}:

\hfill

\theorem \label{_VHS_equi_Intro_Theorem_} 
The VHS $\c H$ is $G_0(M)$-equivariant, under the natural
action of $G_0(M)$ on $C$ and $\c H$.\footnote{The action of $G_0(M)$ on
the trivial bundle $\c H= H^*(M)\times C$ comes 
from a natural action of $G_0(M)$ on the space $H^*(M)$.}

{\bf Proof:} This is an immediate 
implication of \ref{_eta_for_diff_c_H_equiv_Claim_}.
$\;\;\blacksquare$ 

\hfill

To make this statement more explicit, we recall that a variation
of Hodge structures is a flat bundle over a complex manifold, 
equipped with a real structure
and a holomorphic filtration (Hodge filtration), 
which is complementary to its complex conjugate filtration,
and satisfies so-called {\bf Griffiths transversality condition}.
Then, \ref{_VHS_equi_Intro_Theorem_} says 
that the action of $G_0(M)$ on $\c H$ 
maps flat sections to flat sections, and preserves the real structure 
and the Hodge filtration. 

\subsection{On the equivariance of the Yukawa multiplication.}

Let $\c H_{gr}$ be associated graded bundle of the VHS $\c H$,
\[ \c H_{gr}\restrict{P_c(I)} = \oplus H^{p,q}_I(M). \]
 Let $K$ be the pullback of $\calo(1)$ from $\Bbb P H^2(M, \C)$
to $C\hookrightarrow \Bbb P H^2(M, \C)$. Then Yukawa multiplication
on $\c H$ is a map $\c H_{gr} \times \c H_{gr} \arrow \c H_{gr} \otimes K$, 
defined in the same way as the usual Yukawa product for the VHS associated
with Calabi-Yau manifolds (see Section \ref{_VFA_Yukawa_Section_}):

\[ H^{p,q}_I(M) \times H^{p',q'}_I(M) 
   \arrow H^{p+p'-n,q+q'}_I(M)\otimes K\restrict{I}. 
\]
Since the VHS $\c H$ is equivariant, the bundle $\c H_{gr}$
is equipped with a natural $G_0(M)$-equivariant structure. 
The bundle $K$ is also naturally $G_0(M)$-equivariant.
The key theorem of this paper is the following.

\hfill

\theorem \label{_Yu_equi_from_VHS_Theorem_} 
The Yukawa product $\c H_{gr} \times \c H_{gr} \arrow \c H_{gr} \otimes K$
is compatible with the $G_0(M)$-equivariant structure in 
$\c H_{gr}$, $K$. 

{\bf Proof:} This is \ref{_Yu_equiv_Theorem_} .
$\;\;\blacksquare$

\hfill

\ref{_Yu_equi_from_VHS_Theorem_} explains how the Yukawa product varies
with the variation of $x\in Comp$. 

\hfill

\ref{_Yu_equi_from_VHS_Theorem_} is proved by the following argument. 
Let $n = \dim_\C M$. The holomorphic symplectic form $\Omega$
defines an identification 
\begin{equation} \label{_Omega^i_iso_its_dual_Equation_}
   \left(\Omega^i(M)\right)^* \cong \Omega^i(M).
\end{equation}
The top exterior power of $\Omega$ is a non-degenerate section 
of the canonical class $\Omega^n(M)$, because $\Omega$
is symplectic. This section produces an isomorphism
\begin{equation} \label{_Omega^i_iso_Omega^n-i_Equation_}
   \left(\Omega^i(M)\right)^* \cong \Omega^{n-i}(M).
\end{equation}
the composition of \eqref{_Omega^i_iso_its_dual_Equation_}
and \eqref{_Omega^i_iso_Omega^n-i_Equation_} gives an isomorphism
\[
  \eta:\; \Omega^i(M) \arrow \Omega^{n-i}(M)\, .
\]
Now, $\eta$ induces a natural isomorphism of linear spaces,
\begin{equation}\label{_eta_on_coho_Introdu_Equation_}
  \eta:\; H^i(\Omega^j(M)) \arrow  H^i(\Omega^{n-j}(M).
\end{equation}
We call the map \eqref{_eta_on_coho_Introdu_Equation_}
{\bf the Serre duality operator} (Section 
\ref{_operator_Serre_dua_Section_}). By definition,
the Yukawa product in cohomology of $M$ 
coincides with the usual cup-product in
cohomology twisted by $\eta$. 

\hfill

Let $\g(M)\subset End(H^*(M))$ be the Lie algebra 
generated by the Hodge operators
$L_\omega$, $\Lambda_\omega$, where $\omega$ runs through
K\"ahler classes corresponding to all complex structures on $M$.
Let $G(M)\subset End(H^*(M))$ be the Lie group associated with $\g(M)$.
We prove that $\eta$ belongs to $G(M)$ 
(\ref{_Serre_dua_through_g(H)_Theorem_}) and express $\eta$
algebraically in terms of the holomorphic symplectic form
(\ref{_Theta_inde_algebraic_Lemma_}). 

\hfill

In \cite{_main_}, we construct 
$G_0(M))\subset End(H^*(M))$ as a subgroup of
$G(M)\subset End(H^*(M))$. This gives a way to work with $\eta$
in terms of the $G_0(M)$-action. In particular, we obtain the
following theorem, with easily implies \ref{_Yu_equi_from_VHS_Theorem_}.

\hfill

\theorem \label{_G_0_acts_on_Omega_same_on_eta_Theorem_} 
Let $(I, \Omega)$, $(I', \Omega')$ be holomorphic 
symplectic structures on $M$, and $[\Omega], [\Omega']\in H^2(M,\C)$
be the corresponding cohomology classes. Let $g\in G_0(M)$ be a group
element such that, under the natural action of $G_0(M)$ on $H^2(M)$,
$g([\Omega]) = [\Omega']$. Let $\eta$, $\eta'\in End(H^*(M))$ be the
Serre duality operators associated with $(I, \Omega)$, $(I', \Omega')$.
Then

\[ g \eta g^{-1} = \eta'. \]

{\bf Proof:} This is \ref{_eta_for_diff_c_H_equiv_Claim_}. 
$\;\;\blacksquare$

\subsection{On the Tian-Todorov coordinates.}

Let $Comp$ be the (coarse, marked) moduli space of complex structures
on a Calabi--Yau manifold $M$.
The Bogomolov--Tian--Todorov theorem provides 
canonical flat coordinates on $Comp$, the so-called Tian-Todorov coordinates.
We define these coordinates in Section \ref{_Tia_Todoro_coordi_Section_}. 
When $M$ is a holomorphically symplectic manifold, it is possible to
compute the Tian-Todorov coordinates explicitly. 

The moduli space $Comp$ is equipped with a period map 
\[ Comp \arrow {\Bbb P}H^2(M,\C),\] which associates 
a line $l\in H^2(M,\C)$, $l = H^{2,0}_I(M)$ to a complex
structure $I\in Comp$. Let $C$ be the set of all lines satisfying
$(l,l)_{\c H}=0$. From the definition of the form $(\cdot,\cdot)_{\c H}$
it follows that for all $I$, the periods of $L$ lie in $C$.
Let $P_c:\; Comp \arrow C$ be period map.
By Bogomolov and Beauville (\cite{_Beauville_}), 
the map $P_c$ is \'etale. Thus, constructing local coordinates on $Comp$
is equivalent to constructing local coordinates on $C$.

In \cite{_main_}, we proved that the space $H^*(M)$ is equipped with
a natural action of the Lie algebra 
$\g_0(M) \cong \goth{so}(H^2(M, \R), \; (\cdot,\cdot)_{\c H})$.
Let $G_0(M)\subset End(H^*(M))$ be the corresponding Lie group.
We have shown (\cite{_main_}, Corollary 12.5) that $G_0(M)$
acts on on the cohomology ring $H^*(M)$ by automorphisms. 
Clearly, $G_0(M)\otimes \C$ acts naturally on $C$. For
$I\in Comp$, we denote by $ad\,I$ the endomorphism of $H^*(M)$  
defined by $ad\,I (\omega^{p,q}) = \1(p-q) \omega^{p,q}$, 
$\omega^{p,q}\in H^{p,q}(M)$. In \cite{_main_}, we show that
$ad\,I$ belongs to $\g_0(M)\subset End(H^*(M))$ (\cite{_main_},
Theorem 12.2).

Every complex structure $I\in Comp$ defines a decomposition of the
Lie algebra 
\[ \g_0(M)= \g^{I,-2}_0(M) \oplus \g^{I,0}_0(M)
   \oplus\g^{I,2}_0(M)
\]
with 
\[
   \g^{I,i}_0(M) = 
   \{ x\in \g_0(M) \;\; |\;\; [ad\,I, x ] =i \1 x \}. 
\]
Let $G_0^{I, i}(M)\subset End(H^*(M))$ be the Lie group 
associated with $\g^{I,i}_0(M)$.  Then $G_0^{I, 0}(M)$
and $G_0^{I, 2}(M)$ stabilize $P_c(I) \in C$. 
Let $\phi:\; G_0^{I, -2}(M)\arrow C$ map $\alpha \in G_0^{I, 2}(M)$
to $\alpha(P_c(I)) \in C$. The group $G_0(M)\otimes \C$ transitively
acts on $C$. Comparing dimensions of $C$ and $G_0^{I, -2}(M)$,
we find that $\phi$ is an isomorphism locally in a neighbourhood of 
the identity. Since the group $G_0^{I, -2}(M)$ is abelian, 
$\phi$ defines local flat coordinates in $C$.
\footnote{Throughout this paper, the flat coordinates
in a neighbourhood $U$ of $x\in X$
are understood as a set $S$ of pairwise commuting linearly
independent holomorphic vector fields, $|S|= \dim_\C X$, defined in
$U$. We say that $X$ is {\bf equipped with
flat coordinates} if each point $x$ of $X$ has a neighbourhood 
equipped with such set $S_x$ of holomorphic vector fields.
No conditions of compatibility between $S_x$, for different $x$,
is required.}
 Since $P_c$ is \'etale,
$\phi$ also gives local coordinates on $Comp$. In Section 
\ref{_Periods_and_Tia_Todo_coordi_Section_}, we prove that
these coordinates coincide with the Tian-Todorov coordinates.

\hfill

\tableofcontents

\hfill

\hfill

\hrule

\hfill

\begin{itemize}

\item The Introduction (Section \ref{_Intro_Section_}) 
is included to give an outline of our
reasoning. Section \ref{_Intro_Section_} refers to the body
of the article for proofs, but otherwise, Section 
\ref{_Intro_Section_} and the rest of the paper are completely 
independent.

\item Section \ref{_VFA_Yukawa_Section_} gives a number of definitions
culminating in the definition of variation of Frobenius algebras (VFA).
We give an example of VFA (so-called {\bf Yukawa VFA}), which is associated
with every Calabi-Yau manifold. 

\item Section \ref{_Qua_VFA_Section_} provides another example of 
variation of Frobenius algebras
(so-called {\bf Quantum Cohomology VFA}), conjecturally associated with 
every Calabi-Yau manifold. The existence of Quantum Cohomology VFA
is a prerequisite for existence of the Mirror Dual manifold. 
The existence of Quantum Cohomology VFA is proven for 
holomorphically symplectic manifolds of generic type.
In Section \ref{_Qua_VFA_Section_} we follow 
\cite{_Kontsevich-Manin_}, \cite{_Kontsevich_Zurich_}.

\item Section \ref{_Tia_Todoro_coordi_Section_}
gives proofs of 
a number of basic results on deformations of Calabi-Yau manifolds,
mostly due to Bogomolov, Tian and Todorov (\cite{_Bogomolov:78_}
\cite{_Tian_}, \cite{_Todorov:Tia-Todo_}).
Section \ref{_Tia_Todoro_coordi_Section_}
is independent of Sections \ref{_VFA_Yukawa_Section_}--
\ref{_Qua_VFA_Section_}, and uses 
the basics of algebraic geometry and deformation theory 
(\cite{_Griffi_Harri_}, \cite{_Kodaira_Spencer_}).
As a final result, we obtain a definition of canonical 
flat coordinates on the moduli space of Calabi-Yau manifolds.
These coordinates were introduced to Mirror Symmetry by 
\cite{_BerCecOogVaf_}. The exposition follows \cite{_Todorov:Tia-Todo_}.

\item Section \ref{_Mirro_conj_Section_} gives a statement of the
Mirror Conjecture following \cite{_Kontsevich_Zurich_}, 
\linebreak
\cite{_BerCecOogVaf_}.
There are no original results in Sections \ref{_VFA_Yukawa_Section_}
-- \ref{_Mirro_conj_Section_}. Also, nothing in Sections
\ref{_VFA_Yukawa_Section_} -- \ref{_Mirro_conj_Section_}
is in any way specific to hyperk\"ahler or holomorphically
symplectic geometry. We are dealing with manifolds
of hyperk\"ahler type only starting from 
Section \ref{_quantu_coho_for_holo_symple_Section_}.

\item Section \ref{_quantu_coho_for_holo_symple_Section_}
gives an explicit description of Quantum VFA for a holomorphically
symplectic manifold which is generic in its deformation class.
In \cite{Verbitsky:Symplectic_II_}, we proved that 
such manifolds have no rational curves. Since the
``instanton corrections'' (non-trivial terms of the Quantum
product on cohomology) are expressed by counting rational curves, 
it is easy to give an explicit description of Quantum VFA
when all rational curves are trivial. 
 
\item Section \ref{_g(M)_Section_} 
gives an outline of our description of the Yukawa VFA $\underline{\c A}$
for a holomorphically symplectic manifold, with most proofs
postponed till  Section \ref{_equiv_proofs_Section_}.
Let $C$ be the period space of $M$, and $Comp$ be the
(coarse, marked) moduli space of $M$.
By Bogomolov, the period map $Comp \arrow C$ is \'etale.
In  Section \ref{_g(M)_Section_} we explain how $\underline{\c A}$ 
is related to the period space $C$. We identify $C$ 
with a certain orbit in the adjoint representation of the group $SO(m-3,3)$,
where $m= h^2(M)$. We then construct an $SO(m-3,3)$-equivariant VFA $\c A$
on $C$ (existing modulo \ref{_stabi_I_commu_with_Yuka_Theorem_},
which is proven in Section \ref{_equiv_proofs_Section_}). Finally
\ref{_Yu_equiv_Theorem_}, proven in Section \ref{_equiv_proofs_Section_},
states that $\underline{\c A}$ is a pullback of $\c A$ under the
period map. 

\item Sections \ref{_Spin(5)_action_Section_} 
-- \ref{_equiv_proofs_Section_} are dedicated
to the proof of results outlined in Section \ref{_g(M)_Section_}.
These sections are independent of the first part of this paper,
but rely heavily on \cite{_main_} .

\item Section \ref{_Spin(5)_action_Section_} deals with a linear-algebraic
structure of the exterior algebra $\Lambda^*(T)$ 
of a quaternionic-Hermitian space $T$. We explain how 
results about $\Lambda^*(T)$ imply statements about the 
cohomology of a hyperk\"ahler manifold. We construct an action of
the group $Spin(4,1)$ on the cohomology of a hyperk\"ahler manifold
and explicitly describe the action of its center.

\item Section \ref{_operator_Serre_dua_Section_} uses results of
Section 
\ref{_Spin(5)_action_Section_} to describe the operator of Serre duality 
$\eta$. We prove that $\eta$ belongs to the group $Spin(4,1)$ acting on 
the cohomology of $M$. The operator $\eta$ is expressed explicitly via
the natural realization of $Spin(4,1)= Sp(1,1)$ in $End_{\Bbb H}({\Bbb H}^2)$

\item In Section \ref{_equiv_proofs_Section_} we apply the results
of Section \ref{_operator_Serre_dua_Section_} to prove
\ref{_stabi_I_commu_with_Yuka_Theorem_}, \ref{_Yu_equiv_Theorem_}.
We prove the key results about the equivariant structure on the
variations of Hodge structures corresponding to the cohomology
of a holomorphically symplectic manifold.

\item Section \ref{_Periods_and_Tia_Todo_coordi_Section_} computes 
the Tian-Todorov coordinates on $Comp$ 
in terms of the period map, for a holomorphically
symplectic manifold. We use the results of \cite{_main_}, and the
exact form of Tian-Todorov coordinates as given in Section 
\ref{_Tia_Todoro_coordi_Section_}.

\item Section \ref{_proof_mirro_Section_} gives a proof of the Mirror
Conjecture for holomorphically symplectic manifolds. We realy heavily
on results obtained in Sections \ref{_quantu_coho_for_holo_symple_Section_}
-- \ref{_Periods_and_Tia_Todo_coordi_Section_}.

\end{itemize}


\section{Variations of Frobenius algebras.}
\label{_VFA_Yukawa_Section_}


\definition \label{_Fro_alge_Definition_} 
(Frobenius algebras) \ \ 
A Frobenius algebra over a base field $k$
is  a $k$-linear space equipped with a structure of
associative algebra and a linear map $\epsilon:\; A \arrow k$,
such that the bilinear form 
$(\cdot,\cdot)_A:\; a,b \arrow \epsilon (ab)$ is
non-degenerate. Let $(\cdot,\cdot,\cdot)_A:\; a,b,c \arrow \epsilon (abc)$
be thr trilinear form associated with the multiplicative structure in $A$.
The pair $((\cdot,\cdot)_A,\;(\cdot,\cdot,\cdot)_A)$
is called a Frobenius structure on $A$.
Clearly, the knowledge of $((\cdot,\cdot)_A,\;(\cdot,\cdot,\cdot)_A)$ 
suffices to recover the product in $A$.
The linear map $\epsilon:\; A \arrow k$ is often called
{\bf the trace form} of $A$. The Frobenius algebra $A$ is called
{\bf graded} if $A$ is equipped with a grading, which is respected
by the multiplication, and there is a number $n$ such that
$A_n \cong k$ and the map $\epsilon$ factors through the
natural projection $A \arrow A_n$.

\hfill

\definition \label{_weak_VFA_Definition_} 
Let $X$ be a complex variety and $B=B_0 \oplus B_1 \oplus ... \oplus B_n$ be 
a graded holomorphic vector bundle over $X$, equipped
with a non-degenerate holomorphic pairing 
$(\cdot,\cdot):\; B\times B\arrow \calo_X$
and a holomorphic 3-form 
$(\cdot,\cdot,\cdot):\; B\times B\times B\arrow \calo_X$.
Assume that $(\cdot,\cdot),\;(\cdot,\cdot,\cdot)$ define a 
structure of graded Frobenius algebra 
on the fibers of $B$ in every point of $X$. 
Let $\phi:\; TX \hookrightarrow B$ be a morphism
of holomorphic vector bundles.
Then $\bigg( B, \phi$, $(\cdot,\cdot),\;(\cdot,\cdot,\cdot)\bigg)$ 
is called {\bf a weak variation of Frobenius algebras
over $X$}, or simply {\bf weak VFA.} 

\hfill

\definition \label{_CVHS_Definition_} 
A weak complex variation of Hodge structures (or, weak $\C$-VHS)
is the following collection of data.

\begin{description}

\item[1.] A vector bundle $B$ over a complex manifold $X$.

\item[2.] A flat connection $\nabla:\; B \arrow B \otimes \Lambda^1(X)$
on $B$. As usual, the flat connection induces a holomorphic structure
on $B$.

\item[3.] A system of holomorphic subbundles $B^0\subset B^1 \subset
... \subset B_n =B$, which satisfies
$\nabla B_i \subset B_{i+1}\otimes \Lambda^1 X$.

\item[] The filtration $B^0\subset B^1 \subset
... \subset B_n =B$ is called {\bf the Hodge filtration} of the weak $\C$-VHS 
$B$.

\end{description}

\hfill

\definition \label{_Higgs_Definition_} 
Let $(B, \nabla, B^0\subset B^1 \subset
... \subset B_n =B)$ be a weak $\C$-VHS over a complex manifold $X$.
On the associated graded factors, connection $\nabla$ induces
a holomorphic map

\begin{equation}\label{_KS_Definition_}
   KS:\; B_i / B_{i-1} \arrow 
   (B_{i+1}/ B_i)\otimes_{\calo_X} \Omega^1X.
\end{equation}
This map is called {\bf the Kodaira-Spencer map associated with
the weak $\C$-VHS}. The map \eqref{_KS_Definition_} induces a holomorphic 
map 

\begin{equation}\label{_Higgs_definition_Equation_}
   B_{gr} \arrow B_{gr} \otimes_{\calo_X} \Omega^1X, 
\end{equation}
where $B_{gr} = \oplus B_i / B_{i-1} $ is the associated graded
quotient of $B$. The map \eqref{_Higgs_definition_Equation_} is called
{\bf the Higgs field}, or {\bf the Kodaira-Spencer map} associated with 
the weak $\C$-VHS.

\hfill

\definition \label{_VFA_Definition_} 
Let $X$ be a complex manifold. A variation of Frobenius 
algebras (VFA) on $X$ is a weak
$C$-VHS $(B, \nabla, B^0\subset B^1 \subset
... \subset B^n =B)$, equipped with the following data.

\begin{description}

\item[0.] A flat decomposition $B = B^\odd \oplus B^\even$, which
is compatible with the weak $\C$-VHS structure. Let 
\begin{equation*}
\begin{split}
   (B^0)^\even\subset (B^1)^\even \subset
... \subset (B^n)^\even &=B^\even,\\[2mm] 
   (B^0)^\odd\subset (B^1)^\odd \subset
... \subset (B^n)^\odd &=B^\odd, 
\end{split}
\end{equation*}
be the Hodge filtration on $B^\even$,
$B^\odd$.
We denote $(B^i)^\even/(B^{i-1})^\even$ by $A_{2i}$, 
$(B^i)^\odd/(B^{i-1})^\odd$ by $A_{2i+1}$. The bundle
$A=\oplus A_i$ is naturally isomorphic to $B_{gr}$.

\item[1.] An isomorphism $\calo_X \cong (B^0)^\even$.

\item[2.] A structure of weak VFA on $A=\oplus A_i$,
which satisfies the following condition $\bullet$.
Let $\tau:\; TX \arrow A$ be the
map associated with the weak VFA structure,
$\underline t: \; A\otimes TX \arrow A$ be the homomorphism
mapping $a\otimes \overrightarrow v$ to $a\cdot \tau(\overrightarrow v)$ and 
$t:\; A \arrow A\otimes \Omega^1X$ the map obtained by the
duality $(TX)^* \cong \Omega^1X$. 

\item[$\bullet$] Under the natural identification
$A \cong B_{gr}$, the map $t$ corresponds to the Higgs field
(a. k. a. Kodaira-Spencer map).

\end{description}

\hfill

\example 
(Yukawa VFA)
\ \ Let $M$ be a Calabi-Yau manifold, $X$ its (coarse, marked) moduli 
space, and $K$ be the standard line bundle on $X$ with 
$K\restrict I = H^{n,0}_I(M)$
(sometimes called {\bf the determinant bundle}). Let 
$\theta:\; O_U \tilde{\arrow} K$ be a holomorphic trivialization of $K$
over an open set $U\hookrightarrow X$. 
Let $B$ be a VHS over $X$ 
corresponding to the total cohomology space of $M$,
and $B^{gr}$ be the associated graded vector bundle. 
To introduce a VFA on $B$, we need to define an algebraic
structure on $B^{gr}$. 
Let $n=\dim_\C M$. 
The trivialization $\theta:\; O_U \tilde{\arrow} K$ identifies
the sheaf of holomorphic $i$-forms $\Omega^i(M)$ with
$\Lambda^{n-i} TM$, where $TM$ is the holomorphic 
tangent bundle (see Section \ref{_Yukawa_map_Subsection_} 
for details of this identification). Since 
$H^{p,q}(M)= H^q(\Omega^p(M))$,
the isomorphism $\Omega^i(M)\cong\Lambda^{n-i} TM$
results in a natural isomorphism  
$H^{i}(\Omega^{n-i}(M))\cong H^i(\Lambda^{i} TM)$.
The direct sum $\oplus_i H^i(\Lambda^{i} TM)$ is equipped with a natural
multiplicative structure. This gives a multiplication on 
$B= \oplus H^{n-i,i}(M)=\oplus H^{i}(\Omega^{n-i}(M))$. 
The 2-form $(\cdot,\cdot)$ on $B$ comes from the Poincar\'e pairing.
Finally, by Tian-Todorov, $T_IX$ is canonically isomorphic
to $H^1(TM_I)\cong H^{n-1,1}_I(M)$. This gives an embedding
$TM \hookrightarrow B$. Compatibility of these data
and VHS (condition $\bullet$ of \ref{_VFA_Definition_})
is a standard result which 
follows from Kodaira-Spencer theory.
The obtained VFA is called {\bf the Yukawa
VFA associated with the trivialization $\theta:\; O_U \tilde{\arrow} K$.}


\section{Dubrovin algebras and quantum VFA.}
\label{_Qua_VFA_Section_}


In this section, we construct a Mirror counterpart to the
Yukawa variation of Frobenius algebras, which is called 
{\bf quantum VFA}. For a graded linear space $A$, let $\mbox{Aff}(A)$
be an affine supermanifold which corresponds to $A$.

\hfill

\definition \label{_Dubrovin_alge_Definition_} 
\ \ \cite{_Kontsevich-Manin_}\ \  Let $A$ be a graded 
linear space,\footnote{In applications, $A$ is 
a space underlying the cohomology algebra}
and $S= \mbox{Aff}(A)$.
Let $g_{ef}$ be a standard pairing on $A$ considered
as a constant bilinear form on $S$ and
$v_0$ be an even vector field on $S$ (called {\bf the unit field}) 
in $A$. Let $\Phi$ be a function on $S$ 
(defined on all of $S$ or, as in many geometrical cases,
on its open subset) which satisfies
the following assumptions.

\begin{description}

\item[(i)] $v_0(\Phi) = \Phi$ 

\item[(ii)] \ \ 
$\displaystyle \sum_{e,f} \frac {\6^3\Phi}{\6_a \6_b \6_e} g^{ef}
\frac {\6^3\Phi}{\6_f \6_c \6_d} =
(-1)^{\deg a (\deg b + \deg c)}
\sum_{e,f} \frac {\6^3\Phi}{\6_b \6_c \6_e} g^{ef}
\frac {\6^3\Phi}{\6_f \6_d \6_a}$. 

\item[(iii)] For all $s\in S$, the vector $v_0\restrict{s}\in T_s S$
is a unit in the algebra $T_sS$ defined by {\bf (ii)}.

\end{description}

The pair $(A,\Phi)$ is called {\bf a Dubrovin algebra,}
and a function $\Phi$ is called {\bf Dubrovin potential}.

\hfill

Dubrovin algebras have the following properties.

\hfill

\proposition \label{_Dubro_propert_Proposition_} 
\cite{_Kontsevich-Manin_} \ 
Let $\nabla_0$ be a trivial flat connection on
the tangent bundle to $S$. Let $C^a_{b,c}\in S^2\Lambda^1 S\otimes  TS$ 
be the tensor obtained from
$\frac {\6^3\Phi}{\6_c \6_b \6_e}$ by pairing with $g$:

\[ 
  C^a_{b,c} = \sum_{c,d}\frac {\6^3\Phi}{\6_a \6_b \6_d} g_{d,c} 
\]
We consider $C^a_{b,c}$ as 1-form with coefficients in $End(TS)$.
Then

(i) For all $t$, the operator $\nabla_m =\nabla_0 + 
C^a_{b,c}$ is a flat connection
on the tangent bundle to $S$.

(ii) For all $s\in S$, the tensor $C^a_{b,c}$ and the form $g_{e,f}$
define a structure of Frobenius algebra on $T_s S$. 

$\blacksquare$

\hfill

Let $V$ be a compact algebraic (or symplectic, or K\"ahler)
manifold.
In \cite{_Kontsevich-Manin_},%
\footnote{In this account, we follow 
\cite{_Kontsevich_Zurich_}, which differs from
\cite{_Kontsevich-Manin_} in details.}
 Kontsevich and Manin 
define, axiomatically, systems of classes 

\[ I_{g,n;\beta} \in H_*(\bar M_{g,n})
   \otimes \left( H_*(V)^{\otimes n}\right),
\]
where $\bar M_{g,n}$ is the Deligne-Mumford compactification of the
space of curves of genus $g$ with $n$ marked points.
These homology classes are 
called {\bf Gromov-Witten classes}. For a system of
Gromov-Witten classes, Kontsevich--Manin write down 
a power series 
\begin{equation}\label{_pote_by_Gro-Wi_Equation}
   \Phi_\omega(\gamma):=\sum\limits_{\beta\in H_2(V,\Z)} e^{-\int_{\beta}
   \omega}\sum\limits_{n\ge 3}
   \frac{1}{n!}\int\limits_{I_{0,n;\beta}}1_{\overline{\cal
   M}_{0,n}}\otimes\gamma\otimes \dots\otimes\gamma\,.
\end{equation}
depending on a parameter $\omega\in H^{1,1}(V)$, with an argument 
$\gamma\in H^*(M)$(see 
\cite{_Kontsevich-Manin_} for details of this definition).
If this series converges, \cite{_Kontsevich-Manin_}
prove that it converges to a Dubrovin
potential for a Frobenius algebra $A= \oplus H^{p,p}(V)$. The 
Gromov-Witten classes are proven to exist when $V$ 
is a compact algebraic manifold (\cite{_Kontsevich_Zurich_}).
This statement automatically carries over to K\"ahler manifolds due to
results about compactness of Chow schemes for K\"ahler manifolds (see, e. g., 
\cite{_Lieberman_}).%
\footnote{However, in the case we are considering throughout this paper,
$V$ is a holomorphically symplectic manifold without holomorphic
curves, so Gromov-Witten classes are tautological, and 
\eqref{_pote_by_Gro-Wi_Equation} converges automatically
for all $\gamma$.}
The resulting system of Gromov-Witten classes coincides with the 
algebraic one when $V$ is algebraic.

The power series \eqref{_pote_by_Gro-Wi_Equation}
is known to converge when $V$ is rational, and
conjecturally also converges (at least in some 
subset of $S$) for $V$ Calabi-Yau. This convergence
is a part of the Mirror Conjecture. The Dubrovin
potential $\Phi_\omega(\gamma)$ obtained this way is called
{\bf the quantum cohomology potential}. Let $U\subset H^{1,1}(V)$ 
be the domain of convergence for 
$\Phi_\omega\restrict{\goth B\times H^{1,1}(V)}$ considered as a
function of $\omega$, with $\gamma$ running through an infinitesimal 
ball $\goth B$ in $H^*(V)$ with center in $0$.

\hfill

\definition \label{_quantu-coho-Fro_Definition_} 
Consider the 
Frobenius algebra structure defined on the space $H^*(V)$
by the 3-form $\frac{\6^3\Phi_\omega}{\6_a \6_b \6_e}\restrict{0}$,
and the 2-form which is Poincar\'e pairing. Thus obtained
Frobenius algebra is called {\bf the quantum cohomology ring
of $V$, associated with $\omega\in H^{1,1}(V)$.}

\hfill

Let $\c A$ be a trivial bundle over $U$, with a fiber $H^*(V)$.
Dubrovin potential \eqref{_pote_by_Gro-Wi_Equation} 
defines a structure of Frobenius algebra on $\c A$, 
with multiplication in $\c A\restrict{\omega}$, $\omega\in U$ 
defined by the tensor 
\begin{equation} \label{_qua_multi_tenso_Equation_}
   C_{a,b}^c(\omega) =
   \frac {\6^3\Phi_\omega}{\6_a \6_b \6_e}\restrict 0 g^{ec}, 
\end{equation}
where $g^{ef}\in H^*(V) \otimes H^*(V)$ is the 2-vector defined
by the Poincar\'e pairing. This is just another version of 
\ref{_quantu-coho-Fro_Definition_}.

 Let $\nabla_0$ be a trivial
connection in $\c A$, and $\nabla_m := \nabla_0 + C$ be the  connection
defined by the map $C\restrict{\omega}:\; T_\omega U \arrow End(H^*(V))$,
with $C(t)(\alpha) = C_{a,b}^c(\omega)(t,\alpha)$ for all 
$t\in T_\omega U = H^{1,1}(V)$, 
$\alpha \in H^*(V) = \c A\restrict{\omega}$.%
\footnote{In a more general setting, connections $\nabla_0$,
$\nabla_m$ were considered in \ref{_Dubro_propert_Proposition_};
in particular, \ref{_Dubro_propert_Proposition_} immediately
implies that $\nabla_m$ is flat.} 
Let $\c A_i\subset \c A$ be the constant sub-bundle
with a fiber $H^i(V) \subset H^*(V) = \c A\restrict{\omega}$.
Applying the renumbering procedure of
\ref{_VFA_Definition_} {\bf 0.} backwards, we obtain
a decomposition $\c A = \c A^{\even} \oplus \c A^{\odd}$
and a filtration $\c A^{\even}_0\subset \c A^{\even}_1 \subset ...$,
$\c A^{\odd}_0\subset \c A^{\odd}_1 \subset ...$,
on each of these bundles. This constructed
filtration is obviously constant with respect to $\nabla_0$.
Therefore, $\c A$ is naturally isomorphic to its associated
graded quotient $\c A^{gr}$. We obtain that 
the multiplication
\eqref{_qua_multi_tenso_Equation_} is naturally defined on 
$\c A^{gr}$. We denote it by 
$\bullet:\; \c A^{gr} \times \c A^{gr} \arrow \c A^{gr}$.

\hfill

\proposition \label{_qua_VFA_exi_Proposition_} 
(Kontsevich, \cite{_Kontsevich_Zurich_}) \ 
Let $V$ be a manifold equipped with
a system of Gromov--Witten classes $I_{g,n; \beta}$.
Assume that $c_1(V)=0$ and 
for all $n$, $\beta$, the dimension of
$I_{0,n;\beta}$ is the minimum value predicted by 
the Atiyah-Singer theorem:

\[ 
   \dim_\C I_{0,n;\beta} = n + \dim_{\C} V -3. 
\]
Let $U\subset H^{1,1}(V)$  be a domain of convergence
for \eqref{_qua_multi_tenso_Equation_}, and $\c A$ be a
trivial bundle over $U$ with a fiber $H^*(V)$, equipped
with a decomposition 

\[\c A = \c A^\even\oplus \c A^\odd, \] 
a filtration
\begin{equation} \label{_filtra_in_cA_even,odd_Equation_}
   \c A^{\even}_0\subset \c A^{\even}_1 \subset ...,\ 
   \c A^{\odd}_0\subset \c A^{\odd}_1 \subset ...,
\end{equation}
$\c A^{\even}_k\restrict{\omega} = \bigoplus\limits_{i=0}^k H^{2i}(V)$,
$\c A^{\odd}_k\restrict{\omega} = \bigoplus\limits_{i=0}^k H^{2i+1}(V)$,
and connections $\nabla_0$, $\nabla_m$ as above.
Then 
\begin{equation} \label{_nabla_m_pre_filtra_Equation_}
\begin{split}
   \nabla_m \left(\c A^{\even}_k\right) 
&\subset \c A^{\even}_{k+1}\otimes \Omega^1 U\\[2mm]
   \nabla_m \left(\c A^{\odd}_k\right)  
&\subset \c A^{\odd}_{k+1}\otimes \Omega^1 U
\end{split}
\end{equation}

\hfill

\remark 
From \eqref{_nabla_m_pre_filtra_Equation_} it follows immediately that
the filtration \eqref{_filtra_in_cA_even,odd_Equation_} 
and the connection $\nabla_m$ define on the bundles $\c A^{\even}$, 
$\c A^{\odd}$ a complex variation of Hodge structures.

\hfill

{\bf Proof:} Clearly, 
$\nabla_0 \left(\c A^\epsilon_k\right) \subset \c A^\epsilon_{k}\otimes \Omega^1 U$,
for $\epsilon = \even, \odd$. By definition of $\nabla_m$,
to prove \eqref{_nabla_m_pre_filtra_Equation_}, we need only
to show that 
\begin{equation}\label{_C_A_k_subset_A_k+1_Equation_}
  C \left(\c A^\epsilon_k\right) \subset 
  \c A^\epsilon_{k+1}\otimes \Omega^1 U, \ \epsilon = \even, \odd
\end{equation}
Unraveling the definition of $C$, we obtain that 
\eqref{_C_A_k_subset_A_k+1_Equation_} is implied by
 the following lemma.

\hfill

\lemma \label{_qua_pre_gradi_Equation_}%
Let $c\in H^{1,1}(V)$, $\omega\in U$, and 
\[ 
   \bullet_\omega:\; H^*(V) \times H^*(V) \arrow H^*(V)
\]
be the quantum multiplication on $H^*(M)$ associated with
$\omega$ (see \ref{_quantu-coho-Fro_Definition_}). Then
$c\bullet_\omega \alpha \in H^{i+2}(V)$, for all
$\alpha \in H^{i}(V)$.

\hfill

{\bf Proof:} Let \[ (\cdot,\cdot):\; H^*(V) \times H^*(V) \arrow \C\]
be the Poincar\'e form, and 
\[ (x,y,z)_c:\; H^*(V) \times H^*(V) \times H^*(V)\arrow \C\]
be the form defined by $(x,y,z)_c=(x\bullet_\omega y,z)$.
Let 
\[ 
   J_{g,n;\beta}\in \left(H^*(V,{\Bbb Q})\right)^{\otimes n} \arrow
   H^*(\bar M_{g,n}, {\Bbb Q})
\] 
be the map defined by the Poincar\'e duality from a cycle
\[ I_{g,n;\beta}\in \left(H_*(V,{\Bbb Q})\right)^{\otimes n} \otimes
   H_*(\bar M_{g,n}, {\Bbb Q}).
\] Since $\bar M_{0,3}$ is 
a point, we have that $H_*(\bar M_{0,3}, {\Bbb Q})= {\Bbb Q}$.
Therefore, we may consider $J_{0,3,\beta}$ as a map from
$\left(H^*(V,{\Bbb Q})\right)^{\otimes 3}$ to $\Bbb Q$.
From \eqref{_pote_by_Gro-Wi_Equation}, we have

\begin{equation} \label{_q_multi_from_GW_Equation_}
  (c\bullet_\omega \alpha,\alpha') 
    = \sum\limits_{\beta\in H^2(V, \Z)} 
    e^{-\int_V\beta} \frac{1}{3!} J_{0,3,\beta} (c,\alpha,\alpha').
\end{equation}
By the assumptions of \ref{_qua_VFA_exi_Proposition_},
$\dim I_{0,3,\beta}= \dim V$. Therefore, 
\[ J_{0,3,\beta} (c,\alpha,\alpha')=0\] unless 
\[ 
  \dim c + \dim \alpha + \dim \alpha'= 2 \dim_\R V.
\]
On the other hand, the Poincar\'e form is non-zero only on 
cocycles of complementary dimension. Therefore, 
$\dim (c \bullet_x \alpha) = \dim \alpha +2$. This proves 
\ref{_qua_VFA_exi_Proposition_}. $\;\;\blacksquare$

\hfill

\definition\label{_Dubro_VFA_Definition_} 
Under the assumptions of \ref{_qua_VFA_exi_Proposition_},
let 
\begin{description}
\item[1.] $\nabla_m$,
\item[2.] $\c A = \c A^\even\oplus \c A^\odd$,
\item[3.] $\c A^{\even}_0\subset \c A^{\even}_1 \subset ...$,
$\c A^{\odd}_0\subset \c A^{\odd}_1 \subset ...$,
\item[4.] $\bullet:\; \c A^{gr} \times \c A^{gr} \arrow \c A^{gr}$
\end{description}
be the structures we defined above. Clearly,
{\bf 1 - 4} define a VFA on $\c A$.
The so-defined VFA is called {\bf the Quantum VFA 
(or Quantum Deformation VFA)} associated to
the system of Gromov--Witten classes on $V$.

\hfill

\remark  
For $V$ a complex manifold with no holomorphic curves,
the assumptions  of \ref{_qua_VFA_exi_Proposition_} are tautologically
satisfied. Therefore, for such $V$ the Quantum Deformation VFA
is defined correctly.


\section{Tian--Todorov coordinates.}
\label{_Tia_Todoro_coordi_Section_}


\renewcommand{\thesection}{\arabic{section}.\arabic{subsection}}
\renewcommand{\thesubsection}{\arabic{section}.\arabic{subsection}}

In this section, we define the canonical coordinates on
the moduli space of Calabi-Yau manifolds (\cite{_BerCecOogVaf_}), 
to use further in the definition of the Mirror Symmetry.

\hfill

\subsection{Yukawa map.} \label{_Yukawa_map_Subsection_}

Let $M$ be an $n$-dimensional Calabi-Yau manifold equipped 
with a Ricci-flat metric and a non-degenerate section  
of the canonical class. Let $\Lambda^i TM$ be the $i$-th exterior power
of the holomorphic tangent bundle to $M$. Clearly, $\Lambda^iTM$ is
dual to the bundle $\Omega^iM$ of holomorphic differential $i$-forms.
On the other hand, a non-degenerate section  $\Omega$
of the canonical class defines a duality between $\Omega^iM$
and $\Omega^{n-i}M$. Thus, we obtain a natural isomorphism
$\iota:\; \Lambda^iTM \arrow \Omega^{n-i}M$. Since $M$ is 
Ricci-flat, $\Omega$ is parallel by Bochner-Lichnerowicz.
 Therefore, the map $\iota$ is compatible
(up to a coefficient) with the natural Hermitian structures on
the bundles $\Lambda^iTM, \Omega^{n-i}M$. The isomorphism $\iota$ induces an 
isomorphism of cohomology 
\[ \eta:\; H^*(\Lambda^iTM) \tilde\arrow H^*(\Omega^{n-i}M)\] 
which is called {\bf the Yukawa map}. Consider the map of differential 
forms with values in vector bundles,

\begin{equation}\label{_Yukawa_on_forms_Equation_}
\tilde \eta:\; \Lambda^*(\Lambda^iTM) \arrow \Lambda^*(\Omega^{n-i}M)
\end{equation}
Since $\eta$ is compatible with the metric, $\eta$ maps harmonic
forms to harmonic forms. Hence, 
the Yukawa map is induced by $\tilde \eta$.
Further on, we don't discriminate between $\eta$ and $\tilde \eta$.

\subsection{Kodaira--Spencer theory.}
\label{_Kodai_Spe_Subsection_}

Let $[\cdot,\cdot]:\; TM \times TM \arrow TM$ be the commutator 
bracket. Consider the corresponding bracket on the cohomology

\[ [\cdot,\cdot]:\; H^1(\Lambda^1TM)\times H^1(\Lambda^1TM) \arrow
   H^2(\Lambda^1TM)
\]
(often called {\bf the Schouten bracket}; see \cite{_Koszul_}).
We have also a multiplication map 

\[ \cup:\; H^1(\Lambda^1TM)\times H^1(\Lambda^1TM) \arrow
   H^2(\Lambda^2TM)
\]
Both of these maps have their counterparts acting
on differential forms:

\begin{equation}\label{_[],cap_on_diff_forms_Equation_}
\begin{split}
 [\cdot,\cdot]:\; \Lambda^{0,1}(\Lambda^1TM)\times \Lambda^{0,1}(\Lambda^1TM) & \arrow
   \Lambda^{0,2}(\Lambda^1TM),\\
 \cup:\; \Lambda^{0,1}(\Lambda^1TM)\times \Lambda^{0,1}(\Lambda^1TM) &\arrow
   \Lambda^{0,2}(\Lambda^2TM)
\end{split}
\end{equation}

\hfill

Deformations of the complex structure operator 
$\bar \6$ are parametrized by $\bar \6_{new} = \bar \6+ A$, where
$A$ is a $(0,1)$-form with coefficients in $TM$. By definition,

\begin{equation}\label{_A_as_diff_o_Equation_}
  \bar \6_{new} f = \bar \6 f + \sum \lambda_i \cdot \overrightarrow {v_i} (f),
\end{equation}
where $A = \sum \lambda_i \overrightarrow {v_i}$, 
$\lambda_i\in \Lambda^{0,1}(M)$, $\overrightarrow {v_i}$ a holomorphic
vector field. Kodaira--Spencer theory for the
deformations of complex structure can be boiled down to the following
statement.

\hfill

\theorem\label{_Kodai_Spencer_Theorem_}%
(Kodaira--Spencer, Kuranishi) \ \ The operator $\bar \6_{new}$
defines a complex structure if and only if
\begin{equation}\label{_Kodai_Spe_Equation_}
[A,A] = \bar \6 A,
\end{equation} 
where 
$\bar \6:\; \Lambda^{0,q}(M, TM)\arrow \Lambda^{0,q+1}(M, TM)$
is the $\bar\6$-operator on the forms with coefficients in $TM$. 
The forms $A_1, A_2 \in \Lambda^{0,1}(M, TM)$ define isomorphic
complex structures if and only if $A_1 - A_2 = \bar \6 x$,
for $x$ a holomorphic vector field on $M$. 

$\blacksquare$

\subsection{Tian--Todorov lemma.} 
\label{_Tian_Todoro_Lemma_Subsection_}

Under the assumptions of Subsection \ref{_Yukawa_map_Subsection_},
it is possible to relate the maps $[\cdot,\cdot]$ and $\cup$ of
\eqref{_[],cap_on_diff_forms_Equation_}. 

\hfill

\lemma \label{_Tian_Todorov_Lemma_} 
(Tian--Todorov) 
\cite{_Tian_}, \cite{_Todorov:Tia-Todo_}. \ \ 
Let 
\[ \eta:\; 
   \Lambda^{0,q}(\Lambda^iTM) \arrow \Lambda^{0,q}(\Omega^{n-i}M)
\]
be the map of \eqref{_Yukawa_on_forms_Equation_}. Then 
\begin{equation} \label{_Tian_Todo_Equation_}
   \eta([\alpha,\beta]) = \6 (\eta(\alpha) \bullet_{{}_Y} \eta(\beta))
   - \6\eta(\alpha)\bullet_{{}_Y} \eta(\beta) -\eta(\alpha)\bullet_{{}_Y} \6\eta(\beta),
\end{equation}
where $\6:\; \Lambda^{0,q}(\Omega^pM) = \Lambda^{p,q}(M) \arrow
\Lambda^{p+1,q}(M) = \Lambda^{0,q}(\Omega^{p+1}M)$
is the standard Dolbeault differential on $(p,q)$-forms,
 $\alpha$, $\beta\in\Lambda^{0,1}(M, TM)$, and 
\[ 
   \bullet_{{}_Y}: \Lambda^{p,q}(M)\times 
   \Lambda^{p',q'}(M)\arrow \Lambda^{p+p'-n,q+q'}(M)
\] 
is the so-called ``Yukawa product'' on forms, 
$a \bullet_{{}_Y} b = \eta\bigg (\eta^{-1}(a)\cup \eta^{-1}(b)\bigg )$

\hfill

{\bf Proof:} Both sides of \eqref{_Tian_Todo_Equation_}
contain differential operators of first order. Therefore it suffices to
prove \eqref{_Tian_Todo_Equation_} for $M= \C^n$ with the flat metric. 
In this case \eqref{_Tian_Todo_Equation_} is proved trivially by 
computation. $\;\;\blacksquare$

\hfill

\subsection{Green operators and the Schouten bracket.}
\label{_Green_ope_Subsection_}

The Tian-Todorov lemma provides a way to  solve the equa\-tion 
\eqref{_Kodai_Spe_Equation_} explicitly, thus proving the Torelli theorem
for Calabi-Yau manifolds (known as Bo\-go\-mo\-lov-Tian-Todorov theorem).

\hfill

Let 

\[ \bullet_{{}_Y}:\; \Lambda^{p,q}(M)\times \Lambda^{p',q'}(M) \arrow
   \Lambda^{p+p'-n, q+q'}(M)
\]
be the multiplication map obtained from 

\[ \cup:\; \Lambda^{0,q}(\Lambda^{n-p} TM) \times 
   \Lambda^{0,q'}(\Lambda^{n-p'} TM) \arrow 
   \Lambda^{0,q+q'}(\Lambda^{2n-p-p'} TM)
\]
by twisting with $\eta$. The map $\bullet_{{}_Y}$ is 
called {\bf the Yukawa product}.

Applying $\eta$ to both sides of \eqref{_Kodai_Spe_Equation_},
and then using the Tian--Todorov lemma, we obtain the equation

\begin{equation} \label{_Koda_Spence_dual_by_Yuka_Equation_}
    \6 (a \bullet_{{}_Y} a) = \bar\6 a,
\end{equation}
where $a=\eta(A)$ is a section of $\Lambda^{n-1,1}(M)$
satisfying $\6 (a) =0$.
This equation is easier to solve than \eqref{_Kodai_Spe_Equation_}.

\hfill

\lemma \label{_d_bar_d_Lemma_} 
($\6\bar\6$-lemma) \cite{_Griffi_Harri_} \ 
Let $\omega\in \Lambda^{p,q}(M)$ be a form which is $\bar\6$-closed
and $\6$-exact. Then $\omega = \6\bar\6 \theta$ for some 
$\theta\in \Lambda^{p-1,q-1}(M)$. 

$\blacksquare$

\hfill

Let $G_{\bar\6}:\; \Lambda^{p,q}(M) \arrow \Lambda^{p,q-1}(M)$
be the Green operator which inverts the differential
operator $\bar\6$. This operator is defined by 
$G_{\bar\6} = G_{\Delta} \circ \bar \6^*$,
where $G_{\Delta}$ is the usual Green operator inverting the
Laplacian.
Consider the standard orthogonal decomposition 

\[ 
   \Lambda^{p,q} = \im \bar \6 \oplus \im \bar\6^* \oplus {\cal H}^{p,q},
\]
where ${\cal H}^{p,q}$ is the space of harmonic $(p,q)$-forms.  
Then $G_{\bar\6}$ satisfies 

\begin{equation}\label{_Green_properties_Equation_}
\begin{split}
  G_{\bar\6}\restrict{\im \bar\6^* }=&0,\\
  G_{\bar\6}\restrict{{\cal H}^{p,q}}=&0,\\
  G_{\bar\6} \circ \bar \6 \restrict{\im \bar\6} = &Id
\end{split}
\end{equation}
By Kodaira, $G_{\bar\6}$ commutes with $\6$:

\[ 
   G_{\bar\6} \circ \6 = - \6 \circ G_{\bar\6}. 
\]
The following lemma is a key statement in constructing
Tian-Todorov coordinates.

\hfill

\lemma\label{_bar_6_a_n_=0_Lemma_} 
Let $a\in {\cal H}^{n-1,1}(M)$ be a harmonic form on $M$.
We define a form $a_k\in \Lambda^{n-1,1}(M)$ recursively
by

\begin{equation*}\left\{
\begin{split}
   a_k &:=  
\sum_{\begin{array}{c}\scriptscriptstyle {i+j=n-1}\\[-.8mm]
\scriptscriptstyle {i,j\geq 0}\end{array}} 
   G_{\bar \6}\6 \left(a_i\bullet_{{}_Y} a_{j} \right), k \geq 1\\
   a_0 &:=a,
\end{split} \right.
\end{equation*}
where 
\[ \bullet_{{}_Y}:\; \Lambda^{n-1,1}(M) \times \Lambda^{n-1,1}(M)
   \arrow \Lambda^{n-2,2}(M)
\] is the Yukawa product.

Then 
\[ \bar \6 \left( 
            \sum_{{i+j=k-1}} 
    \6 (a_i\bullet_{{}_Y} a_{j}) \right) =0
\] 
for all $k>0$.

\hfill

{\bf Proof:} For $\alpha, \beta \in \Lambda^{p,q}(M)$, let
$[\alpha,\beta] := \6 (\alpha\bullet_{{}_Y} \beta)$. This operation
is Yukawa dual to the Schouten bracket of Subsection 
\ref{_Kodai_Spe_Subsection_}.
Since the Schouten bracket commutes with the holomorphic structure operator,
and the Yukawa duality operator $\eta:\; \Omega^i(M) \arrow \Lambda^{n-1}TM$
is holomorphic, we have

\begin{equation} \label{_bar_6_on_[]_Equation_}
 \bar \6 [ x, y ] = [ \bar \6 x, y ] + [ x, \bar \6 y ]. 
\end{equation}
Therefore, 
\begin{equation*}
    \bar \6 [ a_i, a_{j} ]  =
    [\bar \6 a_i, a_{j} ] + 
    [ a_i, \bar \6 a_{j} ]
\end{equation*}
Using induction, we may assume that

\[ \bar \6 
   \sum_{{i+j=p-1}} 
   [  a_i, a_{j} ] = 0. 
\]
for $p<n$. Using the $\6\bar\6$-lemma and \eqref{_bar_6_on_[]_Equation_}, 
we obtain that for such $p$, the form 

\[ \sum_{i+j=p-1}
   [a_i,a_{j}] =  \6 \sum_{{i+j=p-1}}
   a_i\bullet_{{}_Y} a_{j}
\]
is $\bar\6$-exact. Therefore, by \eqref{_Green_properties_Equation_},
we have
\begin{equation} \label{_d_a_i_sum_commutato_Equation_}
   \bar \6 a_p = \bar \6 G_{\bar \6} \left( 
   \sum_{{i+j=p-1}} [a_i,a_j] \right)
    = 
   \sum_{{i+j=p-1}} [a_i,a_j]
        .
\end{equation}
This implies that

\[ \bar\6 [a_p, a_q] =    [\bar \6 a_p, a_{q} ] + 
    [ a_p, \bar \6 a_{q} ] =  \sum_{{i+j=p-1}} [[a_i,a_j], a_q] +
   \sum_{{i+j=q-1}} [a_p, [a_i,a_j]]. 
\]
Therefore, 
\begin{equation} \label{_bar6(commmu)_3-sum_Equation_}
 \bar \6 \sum_{{p+q=k-1}} [a_p,a_q] = 
   \sum_{i+j+l = k-2} \bigg( [[a_i,a_j], a_l] + [a_i,[a_j,a_l]] \bigg) 
\end{equation}
Clearly, the bracket $[\cdot,\cdot]$ is supersymmetric:

\[ [\alpha,\beta] = (-1)^{(p-1)(q-1)} [\beta,\alpha], \ \ 
   \alpha \in \Lambda^{n-p,p}(M), \beta\in \Lambda^{n-q,q}(M)
\]
This implies that the sum \eqref{_bar6(commmu)_3-sum_Equation_}
is zero. \ref{_bar_6_a_n_=0_Lemma_} is proven. $\;\;\blacksquare$

\subsection{Canonical coordinates.}

We use the notation introduced in Subsection \ref{_Green_ope_Subsection_}.
For any $a\in {\cal H}^{n-1,1}(M)$, consider the sum $\sum a_i$,
where the $a_i$ are the forms defined in \ref{_bar_6_a_n_=0_Lemma_}.
Clearly from Hodge theory, the operator $G_{\bar \6}$ is 
compact and $\6$ is elliptic. From spectral theory it might be seen that
the sum $\sum a_i$ converges absolutely for sufficiently small $a$.
Let $A\in \Lambda^{0,1}(TM)$ be the $(0,1)$-form with coefficients in 
$TM$ which corresponds to 
$\sum a_i \in \Lambda^{0,1}\left(\Omega^{n-1}(M)\right)$
by the Yukawa isomorphism \eqref{_Yukawa_on_forms_Equation_}.

\hfill

\claim \label{_A_sati_Kodai_Claim_} 
The form $A$ satisfies the equation \eqref{_Kodai_Spe_Equation_}
of Kodaira-Spencer.

{\bf Proof:} 
By construction, $\6 a_i =0$, for all $i$. Therefore,
by Tian-Todorov, to prove \eqref{_Kodai_Spe_Equation_}
it suffices to show that $A' = \sum a_i$
satisfies $\6(A'\bullet_{{}_Y} A') =\bar\6 A' $.
By definition,
\[\bar\6 A'  = 
  \sum_p \bar\6 G_{\bar \6} \left(\sum_{i+j=p-1} [a_i,a_j]\right). 
\] 

By \eqref{_d_a_i_sum_commutato_Equation_} 
and \ref{_bar_6_a_n_=0_Lemma_}, we have

\[ \bar\6 G_{\bar \6} \sum_{i+j=p-1} [a_i,a_j] = \sum_{i+j=p-1}
   [a_i,a_j]. 
\]
Therefore, 
\[ \bar\6 A'  = \sum_p \sum_{i+j=p-1}
   [a_i,a_j] = [A', A'].
\]
$\;\;\blacksquare$

\hfill

We obtain that $A$ defines a complex structure on $M$,
in the sense of Kuranishi (\ref{_Kodai_Spencer_Theorem_}).
We have proved the following theorem.

\hfill

\theorem \label{_Bogo_Tian_Todo_Theorem_} 
(Bogomolov -- Tian -- Todorov) \ \ \cite{_Bogomolov:78_}
\cite{_Tian_}, \cite{_Todorov:Tia-Todo_} \ \ 
For each cohomology class $\alpha \in H^1(M, TM)$,
consider the corresponding harmonic form $a\in {\cal H}^{n-1,1}(M)$.
Let $B \subset H^1(M, T)$
be an open ball where the series $\sum a_i$ converges.
Let $X$ be the moduli space of $M$. Let 
$\phi:\; B\arrow X$ be the map associating to $\alpha$ 
the complex structure defined by $A = \sum a_i$. Then, for 
$B$ sufficiently small, the map
$\phi$ is an open embedding, which is independent of  
the choice of K\"ahler metric on $M$.

{\bf Proof:} Immediately 
follows from \ref{_Kodai_Spencer_Theorem_}. $\;\;\blacksquare$

\hfill

\definition \label{_Tian_Todo_coord_Definition_} 
\ \ \ Let $M$ be a Calabi-Yau manifold and $x\in X$ be a point in 
its moduli space. Consider the coordinates in a neighbourhood of
$X$, defined as in \ref{_Bogo_Tian_Todo_Theorem_}.
These coordinates are called {\bf Tian-Todorov coordinates on $X$.}

\hfill

\remark 
Tian--Todorov coordinates depend on the base point $x\in X$.
The translation between these coordinates, for different base points,
is not necessarily flat.

\hfill

\remark 
In the context of the 
Mirror Symmetry, Tian--Todorov coordinates were introduced in
\cite{_BerCecOogVaf_}. Since then, these coordinates have been common in
the physics literature. The definition above comes from 
A. Todorov (\cite{_Todorov:Tia-Todo_}; also \cite{_Tian_}).

\renewcommand{\thesection}{\arabic{section}}


\section{The Mirror Conjecture.}
\label{_Mirro_conj_Section_}


\subsection{Mirror Symmetry from a mathematician's point of view.}

\definition\label{_equiv_Yuka_Definition_} 
Let $M$ be a Calabi-Yau manifold, and $X$ be its deformation space. 
The Tian-Todorov lemma provides that for every $I\in X$,
there are canonical flat\footnote{These coordinates depend on $I$, 
in such a way that translation between $U_I$ for different
$I$ is {\it not} flat.} coordinates in a neighbourhood
$U_I$ of $X$ (first introduced in 
\cite{_BerCecOogVaf_}; for  a formal definition,
see Section \ref{_Tia_Todoro_coordi_Section_} 
of the present paper). These coordinates are called {\bf Tian-Todorov 
coordinates on $X$}. Denote by $K$ the 
standard line bundle over $X$, $K\restrict x = H^{n,0}_x(M)$.
Let $\c A$ be a VFA over $U$, where $U$ is an open subset
in a linear space $L$. We say that
{\bf Yukawa VFA is equivalent to $\c A$}
if for all $I\in X$ there exist 

\begin{description}

\item[(i)] An open set $U_0 \subset U$ and an isomorphism 
$\phi_I:\; U_0 \arrow U_I$, preserving flat coordinates,
where $U_I$ is a Tian-Todorov
neighbourhood of $I$ equipped with flat coordinates.

\item[(ii)] A trivialization $\theta$ of $K$ over $U_I \subset Comp$.

\item[] {\small Let $\mbox{Yu}$ be the Yukawa VFA on $U_I$ defined by $\theta$,
and $\c H$ be its total space, $\c H = H^*(M) \times U_I$.} 

\item[(iii)] An isomorphism $\phi_I^*A\sim \c H$ of vector bundles
over $U_I$ inducing an isomorphism of variations of 
Frobenius algebras.\footnote{The VFA structure on $\phi_I^* A$ 
is the pullback of the VFA on $A\restrict{U_0}$.}
\end{description}

\hfill

{\bf Mirror Conjecture:} Let $M$ be a Calabi-Yau manifold.
We say that {\bf the Mirror Conjecture holds for $M$} if 
there exist a Calabi-Yau
manifold $W$ (called {\bf Mirror dual} Calabi-Yau manifold)
such that the power series \eqref{_pote_by_Gro-Wi_Equation}
converges for $M$ and $W$ in nonempty open sets, the assumptions of
\ref{_qua_VFA_exi_Proposition_} are satisfied for $M$ and $W$,  
Yukawa VFA for $M$ is equivalent to
Quantum VFA for $W$, and Yukawa VFA for $W$ is equivalent to
Quantum VFA for $M$. 

\hfill

\remark 
The quantum VFA $Q = \oplus H^{p,q}(M)$ has a subalgebra $Q^{p,p} = \oplus
H^{p,p}$. Similarly, the Yukawa VFA has a subalgebra 
\[ 
    Y^{n-p,p} = \oplus H_I^{n-p,p}(M).
\] 
Often, Mirror Symmetry is understood as an isomorphism 
between these subalgebras. Our proof of Mirror Symmetry
for holomorphically symplectic manifolds works equally well in both 
of these cases.

\hfill

\remark 
It is important to notice that, in our definition,
the Quantum VFA depends on the trivialization $\theta$ 
of the linear bundle $K$. To wit, Mirror Symmetry
gives information about multiplication in the cohomology
(by identifying Yukawa and Quantum rings) but only up to a scalar 
multiple.

\hfill

The following theorem is the main result of this paper.
It is proven in Section \ref{_proof_mirro_Section_}.

\hfill

\theorem \label{_mirror_for_hype_Theorem_} 
Let $M$ be a compact holomorphically symplectic manifold, 
which is generic in its deformation class.\footnote{$Pic(M)=0$ 
will suffice.} Then the Mirror Conjecture holds for $M$,
which is Mirror dual to itself.

\subsection{Appendix: Mirror Symmetry from a physicist's point of view.}

For a physicist, it is more natural to think of Mirror Symmetry in terms
of the so-called correlation functions. For completeness, we give
an (equivalent to ours) statement of the Mirror Conjecture in these terms. 
We assume that $M$ and $W$ are mirror dual and explain what it
means in the language of correlators.

Let \[ x\in H^{1,1}(W), \ \ y\in Comp(M), \ \ y = \phi_I(x),\] where $\phi_I$
is the Tian-Todorov map. Then Yukawa (B-model) correlations are $i$-forms
on $\oplus H^{p,q}_y(M)$ defined by 

\begin{equation} \label{_B_model_correlations_Equation_}
   \langle \alpha_1 ... \alpha_i \rangle \arrow
   \int_M \alpha_1 \bullet_{{}_Y} ... \bullet_{{}_Y} \alpha_i
\end{equation}
where $\bullet_{{}_Y}$ is the Yukawa product associated with the complex
structure $y$. The quantum (A-model) correlations
are, similarly, $i$-forms on $\oplus H^{p,q}(W)$ 
defined by

\begin{equation} \label{_A_model_correlations_Equation_}
   \langle \alpha_1 ... \alpha_i \rangle \arrow
   (\alpha_1 \bullet_Q ... \bullet_Q \alpha_{i-1}, \alpha_i)
\end{equation}
where $\bullet_Q$ is the quantum product associated with
$x\in H^{1,1}(W)$ and $(\cdot,\cdot)$ is the Poincar\'e 
form on the quantum cohomology (which is a part of the Frobenius
structure on quantum cohomology). The {\bf marginal correlators}
are, in the B-model situation, $n$-forms on $T_y Comp = H^{n-1,1}_y(M)$ 
defined by \eqref{_B_model_correlations_Equation_},
and, on the A-model side, $n$-forms on $H^{1,1}(W)$ defined
by \eqref{_A_model_correlations_Equation_}. 
The ``parameter space'' for the A-model is the convergence domain
of the Dubrovin potential in $H^{1,1}(W)$ (or a quotient thereof by
an action of a discrete group, but we are interested only in the
local structure of the parameter space). 
The ``parameter space'' for the A-model is the image of $\phi_I$ in
$Comp(M)$, or, what is equivalent, an open subset in $H^{n-1,1}(M)$.
Denote the parameter spaces of the respective models by $P_A$, $P_B$.
The Mirror Conjecture says that (shrinking $P_A$, $P_B$ if necessary)
under the appropriate flat  identification of
$P_A \subset H^{1,1}(W)$, $P_B \subset H^{n-1,1}(W)$,
$H^{1,1}(M) \cong H^{n-1,1}(M)$, the bundles 
$\oplus H^{p,q}_y(M)$, $\oplus H^{p,q}(W)$ can be identified
in such a way that correlation functions are mapped to correlation
functions (this corresponds to compatibility of $\phi_I$ with
the Frobenius structure), and marginal correlators to marginal 
corelators (this corresponds to compatibility of $\phi_I$
with the embeddings $\tau:\; TU \arrow \c A^{gr}$).
Notice that, for a physicist, only weak VFA are of significance.
Still, it is easier and conceptually more natural
to formulate the Mirror Conjecture using
the VFA and weak $\C$-VHS.


\section{Quantum VFA for holomorphically symplectic manifolds.}
\label{_quantu_coho_for_holo_symple_Section_}


Let $M$ be a compact holomorphically 
symplectic manifold which is
ge\-ne\-ric in its deformation class. In \cite{Verbitsky:Symplectic_II_} 
it is proven that
all closed complex subvarieties of $M$ are even-dimensional.
In particular, $M$ has no curves (and no integral (1,1)-cycles).
Therefore, all non-trivial Gromov-Witten classes for $M$
vanish. Applying the formula \eqref{_pote_by_Gro-Wi_Equation},
we find that the quantum deformation potential 
is \[\Phi_\omega(\gamma)= \epsilon(\exp(x)),\] where
$\epsilon:\; H^*(M)\arrow \C$ is the trace form projecting $H^*(M)$ to
$H^{2n}(M)=\C$, and $\exp:\; H^*(M)\arrow H^*(M)$ maps $x$
to $1+ x+ \frac{x^2}{2!} + \cdots +\frac{x^n}{n!}+\cdots$.
Note that $\Phi_\omega(\gamma)$ is independent of $\omega$.

For an arbitrary supercommutative Frobenius algebra $A$ over $\C$ or $\R$, 
the function $\Phi(x)= \epsilon(\exp(x))$ satisfies the conditions of 
\ref{_Dubrovin_alge_Definition_}. We call it
{\bf the primary Dubrovin potential 
associated with the algebra $A$.} 

\hfill

\proposition\label{_3-form_for_primary_Du_pote_Proposition_} 
Let $A$ be a supercommutative Frobenius algebra 
over a field of char 0, and $\Phi$ be a primary Dubrovin potential.
Then

\[ \frac{\6^3\Phi}{\6_a \6_b \6_c} (x,y,z)\restrict{t}= 
   \epsilon (xyz \exp(t)),
\]
where $a,b,c$ are arbitrary vectors in $T_t S= A$.

{\bf Proof:} Clear. $\;\;\blacksquare$

\hfill

\corollary \label{_multi_for_primary_Du_pote_Proposition_} 
Let $C^c_{a,b}(x):\; A \times A \arrow A$ be the product in $T_x S$ determined
by the tensor $\frac{\6^3\Phi}{\6_a \6_b \6_c}\restrict{x}$ as in 
\ref{_Dubro_propert_Proposition_}. Let $C^c_{a,b}(0)$ be the original
product in $A$. Then

\[ 
   C^c_{a,b}(t) (x,y) = C^c_{a,b}(0) (x,\exp(t)y). 
\]

{\bf Proof:} Follows from \ref{_3-form_for_primary_Du_pote_Proposition_}. 
$\;\;\blacksquare$

\hfill

We obtain the following description of Quantum VFA.

\hfill

\definition \label{_trivi_VFA_Definition_}
Let $A = A_0 \oplus A_1 \oplus ... \oplus A_n$ be a graded
Frobenius algebra and $U \subset \Aff(A_2)$ be a submanifold
of $\Aff(A_2)$. Let $\c A=  U \times A$ be the trivial bundle over $U$,
with fiber $A$, and $\nabla_0$ be the trivial connection
in $\c A$. There is a natural embedding 
$TU\stackrel \tau\hookrightarrow \c A$ and a natural
multiplication $\cdot:\;\c A\times \c A \arrow \c A$.
Let $C^c_{a,b}:\;  \c A \otimes TU \arrow \c A$
be the map associating $\tau(t)\cdot a$ to $a \otimes t$, where
$a\in \c A$, $t\in TU$, and $t\cdot a$ is multiplication 
in $\c A$. Consider $C^c_{a,b}$ as a 
map from $\c A$ to $\c A\otimes \Lambda^1U$. Let $\nabla_m$ be the 
connection in $\c A$ defined by $\nabla_m = \nabla_0+ C^c_{a,b}$.
Consider the decomposition of $\c A$ onto even and odd parts,
 $\c A = \c A^\odd\oplus \c A^\even$ ,
 with Hodge filtration in
$\c A^\odd$, $\c A^\even$ defined as in \ref{_qua_VFA_exi_Proposition_},
$\c A^{\even}_k = \bigoplus\limits_{i=0}^k \c A_{2i}$,
$\c A^{\odd}_k = \bigoplus\limits_{i=0}^k\c A_{2i+1}$.
Since the filtration $\c A_0^\even \subset \c A_1^\even \subset ...,\ \ 
    \c A_0^\odd \subset \c A_1^\odd \subset ...$ 
comes from the grading, we may naturally 
identify $\c A$ with its associated graded 
bundle $\c A^{gr}$.
Clearly, 
\begin{multline*}
\bigg(
    \c A = \c A^\odd\oplus \c A^\even, \nabla_m, 
    \c A_0^\even \subset \c A_1^\even \subset ...,
    \c A_0^\odd \subset \c A_1^\odd \subset ... \\
    \cdot:\;\c A^{gr}\times \c A^{gr} \arrow \c A^{gr}, 
\bigg)
\end{multline*}
gives a VFA structure on $\c A$. Such VFA is called
{\bf the trivial VFA associated with $A$, $U \subset \Aff(A_2)$}.

\hfill

\theorem \label{_QD_for_hype_trivial_Theorem_} 
Let $M$ be a K\"ahler manifold without
holomorphic curves, $A = H^*(M)$, $U = \Aff(H^{1,1}(M))$,
and $\c A$ be the trivial bundle over $U$ with the fiber $H^*(M)$,
equipped with a Quantum VFA structure. Then $\c A$ is the trivial 
VFA over $U$.

{\bf Proof:} Clear. $\;\;\blacksquare$

\hfill

This finishes the calculation of quantum cohomology VFA in the case
of a holomorphically symplectic manifold which is generic in its
deformation class.

\hfill

{\bf Appendix.}
To see how our computation works in physicists' language, 
the reader can check the following trivial claim.
The ``marginal correlations'' of Section \ref{_Mirro_conj_Section_}
are easy to write down. Let $S^{1,1}$ be the base for 
the Quantum VFA, denoted by $\cal A$, and $i:\; TS^{1,1} \hookrightarrow B$
be the homomorphism which is a part of the VFA structure. 
Since the bundle $TS^{1,1}$ is trivial, we may consider
$i$ as a map from the bundle $TS^{1,1}= H^{1,1}(M)\times S^{1,1}$ 
to $\cal A$. By definition, marginal correlations are $n$-linear functions
on $H^{1,1}(M)$, depending on a base point $s\in S^{1,1}$.
From \ref{_QD_for_hype_trivial_Theorem_} 
it follows that the marginal correlation function is a map
from $S^{1,1}$ to the space $S^n(H^{1,1}(M))^*$ of symmetric $n$-forms
on $H^{1,1}(M)$, given by

\[ \langle \alpha_1, ... \alpha_n \rangle_x =
   \bigg( i(\alpha_1) \bullet_x \ ... \ 
      \bullet_x i(\alpha_{n-1}), i(\alpha_n) \bigg)
\]
as in \eqref{_A_model_correlations_Equation_}.

\hfill

\claim \label{_margi_corre_for_hol_sympl_Claim_} 
Let $\cal A$ be the Quantum VFA associated with a 
manifold which has no rational curves.
Then the  ``marginal correlator'' 
map 
\[ 
   \langle \cdot,\cdot, \ ...\  \cdot \rangle_x:\; 
   S^{1,1} \arrow S^n(H^{1,1}(M))^*
\]
is constant.

{\bf Proof:} Clear. $\;\;\blacksquare$


\section[Periods of holomorphically symplectic 
manifolds and Yu\-ka\-wa 
VFA.]{Periods of holomorphically symplectic 
\\manifolds and Yukawa VFA.}
\label{_g(M)_Section_}


For clarifications and missing definitions, the reader is referred
to \cite{_main_}.

\hfill

In this section, we give an outline of how the Yukawa product on 
the cohomology of a holomorphically symplectic manifold $M$ can 
be expressed through a group action on its period space. 

Let $M$ be a compact holomorphically symplectic manifold of K\"ahler type,
and $Comp$ be its coarse marked deformation space. 
Assume\footnote{This assumption is needed only 
to simplify the exposition. Details in \cite{_main_}, Section 3.}
that $\dim H^{2,0}(M) = 1$.
In \cite{_Beauville_} (Remarques, p. 775),
Beauville defined a canonical non-degenerate symmetric $2$-form
$(\cdot,\cdot)_{\c H}$ on $H^2(M)$ of signature $(n-3,3)$, $n= \dim H^2(M)$.
We call this form {\bf the Bogomolov-Beauville pairing}.
For a complex structure $I\in Comp$, let 
$\rho_I:\; \goth{u}(1) \arrow End(H^*(M))$ 
be the linear operator mapping $t\in \goth{u}(1) \equiv \R$ to 
the endomorphism \[ \omega^{p,q} \arrow (p-q)\1\omega^{p,q}t,\] for all
$\omega^{p,q}\in H^{p,q}_I(M)$.
Let $\g_0(M)\subset End(H^*(M))$ be the Lie algebra 
generated by $\rho_I$ for all $I\in Comp$, and 
$G_0(M)\subset End(H^*(M))$ be the corresponding Lie group.
As \cite{_main_}, Theorem 5.1 implies, $\rho_I$ preserves 
the form $(\cdot,\cdot)_{\c H}$,
for all $I\in Comp$. This defines a Lie algebra homomorphism

\begin{equation}\label{_embe_g(M)_to_so_Equation_}
 \rho:\; \g_0(M)\arrow \goth{so}\bigg( H^2(M), (\cdot,\cdot)_{\c H} \bigg).
\end{equation}

\hfill

\theorem \label{_g(M)=so_Theorem_} 
(\cite{_main_}, Theorem 12.2)
The map \eqref{_embe_g(M)_to_so_Equation_} is an isomorphism.

$\;\;\blacksquare$

\hfill

Let $u$ be the unit vector in $\goth{u}(1)$, and 
$ad\,I= \rho_I(u):\; H^2(M) \arrow H^2(M)$ be the corresponding
endomorphism.  Let $X\subset \g_0(M)$ be
the $G_0(M)$-orbit of $ad\,I$, for some $I$. As the following claim
implies, $X$ is independent of $I\in Comp$.

\hfill

\claim\label{_ad_I_in_the_orbit_Claim_} %
For all $L\in Comp$, the endomorphism $ad L\in End(H^*(M))$ belongs
to $X$.

\hfill

{\bf Proof:} By definition, the endomorphism 
$ad L$ belongs to $\g_0(M)$. On the other hand, we proved
that $\g_0(M)$ is naturally isomorphic to $\goth{so}(\Bbb V)$,
where $\Bbb V$ is the space $H^2(M, \R)$ equipped with the 
Bogomolov-Beauville form $(\cdot,\cdot)_{\c H}$. Consider 
$ad L$ as an endomorphism of $\Bbb
V$. The operator $ad L$ has two non-zero eigenvalues, $2\1$
and $-2\1$, corresponding to $H^{2,0}_L(M)$ and $H^{0,2}_L(M)$
respectively. 
Let $V_{\neq 0}$ be the subspace of $V$ corresponding
to these eigenvalues, 
\[ V_{\neq 0}= \left( H^{2,0}_L(M) 
   \oplus H^{0,2}_L(M)\right) \cap H^2(M,\R). 
\]
Since
$h^{2,0}(M) =1$, the space $V_{\neq 0}$ is 2-dimensional.
 Then $(\cdot,\cdot)_{\c H}$ restricted to $V_{\neq 0}$
is positive definite. Now, \ref{_ad_I_in_the_orbit_Claim_} is an
implication of the following linear algebra observation.

\begin{itemize}
\item  For all $x\in \goth{so}(\Bbb V)$ with 
two non-zero eigenvalues $2\1$
and $-2\1$ of multiplicity one, 
and such that $(\cdot,\cdot)_{\c H}$ restricted to $V_{\neq 0}$
is positive definite, we have $x\in X$. 
\end{itemize}
$\blacksquare$

\hfill

Denote the map
$I\arrow ad\,I$ by $P_o :\; Comp \arrow X$.
Identifying $\g_0(M)$ with its dual space, we can consider $X$ as a coadjoint
orbit. Therefore, $X$ is equipped with a natural complex structure
(as a coadjoint orbit of a reductive group).\footnote{
By Bogomolov, the map $P_o$ is
an open covering, and is holomorphic with respect to this complex structure.}

 For $I\in Comp$, let $G^I_0(M)\subset G_0(M)$ be the stabilizer 
of $P_o(I)\in \g_0(M)$, under the adjoint action of $G_0(M)$ on $\g_0(M)$:

\[ 
   G_0^I(M)  = 
   \left\{ g\in G_0(M) \;\; |\;\; g(\rho_I(u)) = ad\,I\right\}. 
\]

Let $Comp$ be the moduli space of $M$, and
$\underline K$ be the standard holomorphic line bundle over 
$Comp$ with fibers
$\underline K \restrict I = H^0_I(\Omega^{n}(M))$, 
$n =\dim_\C M$, $I\in Comp$.
Let $\underline \Omega$ be the standard 
holomorphic line bundle over $Comp$ with fibers
$\underline \Omega \restrict I = H^{2,0}_I(M)$. By definition, 
$\underline K = \underline \Omega^{n/2}$.
Let $\Omega\subset X \times H^2(M, \C)$ be the line 
bundle over $X$, with fibers
\[ \Omega\restrict x = 
   \{ l \in H^2(M, \C) \;\;|\;\; xl = 2\1 l\}. 
\]
Clearly, $\underline \Omega = P_o^* \Omega$. 
Therefore, $P_o^*(\Omega)^{\frac{n}{2}} = \underline K$.
The bundle $\Omega$ is naturally $G_0(M)$-equivariant. This gives
an action of $G_0^I(M)$ on $\Omega\restrict{ad\,I}$, for all $I\in Comp$.
Since the fiber of $\Omega^{n/2}$ in $ad\,I\in X$ is naturally
isomorphic to the fiber of   $\underline K$ in $I\in Comp$, we obtain
also a natural action of $G_0^I(M)$ on 
$\underline K\restrict{I}\cong H_I^{n,0}(M)$.

\hfill

The following theorem will be proven in Section \ref{_equiv_proofs_Section_}.

\hfill

\theorem \label{_stabi_I_commu_with_Yuka_Theorem_} 
{\it For $I\in Comp$, consider the natural
action of the group $G_0^I(M)$ on $H^*(M)$, $H_I^{n,0}(M)$.
Let 

\[ Y_I:\; H^*(M) \times H^*(M) 
   \arrow H^*(M) \otimes H^{n,0}_I(M)
\]
be the Yukawa product.
Then $Y_I$ commutes with the action of $G_0^I(M)$.}

\hfill

Consider the trivial bundle $B$ 
over $X$, with the fiber $H^*(M)$, $B= X \times H^*(M)$. 
The action of $G_0(M)$ on $H^*(M)$, $X$ defines a $G_0(M)$-equivariant
structure on $B$. 
\ref{_stabi_I_commu_with_Yuka_Theorem_} automatically
allows us to define an equivariant
mutiplicative structure $\bullet_B:\; B \times B \arrow B\otimes K$ 
on $B$ which can be, a priori, dependent 
on the choice of $I$. For $x\in X$, 
$h_1, h_2\in B\restrict{x}$, $x = g(P_o(I))$, 
$\lambda\in \underline K\restrict{x}$, we set

\begin{equation} \label{_equiv_multi_defini_Equation_}
    h_1 \bullet_{B,g} h_2\restrict{x} = 
    g(\lambda)g(Y_I(g^{-1} h_1, g^{-1} h_2))\restrict{g(x)}. 
\end{equation}
To prove that \eqref {_equiv_multi_defini_Equation_}
defines an equivariant product $h_1 \bullet_B h_2$ on $B$, 
we have to show 
that this $\bullet_{B,g}$ is independent on the choice of $g$. 
For two possible choices $g$, $g'$, we have $x= g^{-1}g' \in G_0^I(M)$,
because $G_0^I(M)$ is the stabilizer of $P_o(I)$.
Clearly, from the definition, 
\[ h_1 \bullet_{B,g} h_2 = 
   x^{-1}\left( x(h_1) \bullet_{B,g'} x(h_2)\right).
\]
\ref{_stabi_I_commu_with_Yuka_Theorem_} implies that 
$h_1 \bullet_{B,g} h_2= h_1 \bullet_{B,g'} h_2$.

\hfill

\theorem \label{_Yu_equiv_Theorem_} 
{\it Let $P_o^*(B)$ bethea trivial bundle  over $Comp$ with fiber 
$H^*(M)$, and 

\[ \bullet_{{}_Y}:\; P_o^*(B) \times P_o^*(B)  
   \arrow P_o^*(B) \otimes K
\] 
be the Yukawa product in $P_o^*(B)$. Let $\bullet_B$ 
be the equivariant product in $B$ associated to $I\in Comp$ as above. Then
$\bullet_B$ is independent of $I$, and $\bullet_{{}_Y}$ coincides
with the pullback of $\bullet_B$.}

\hfill

\hfill

\ref{_Yu_equiv_Theorem_} will be proven in Section 
\ref{_equiv_proofs_Section_}.


\section{The $Spin(5)$-action on $H^*(M)$.}
\label{_Spin(5)_action_Section_}

\nopagebreak

We refer to \cite{_main_} for details of definitions
and for missing proofs. 
A hyperk\"ahler manifold is a Riemannian manifold 
$M$ equipped with three complex structures $I$, 
$J$ and $K$, such that $I\circ J=-J\circ I=K$ and $M$
is K\"ahler with respect to $I$,
$J$ and $K$. Relations between $I$, $J$ and $K$
imply that there is an action of the quaternions in its tangent
space. 

Let $M$ be a complex manifold which admits a hyperk\"ahler
structure. A simple linear algebra argument implies that
$M$, considered as a complex manifold with the complex structure 
coming from the standard embedding $\C \hookrightarrow {\Bbb H}$, 
is equipped with a holomorphic symplectic form.
\footnote{Precisely, let $(I,J,K)$ be the standard generators of
the quaternion algebra, and $\omega_\alpha$, $\alpha = I,J,K$ K\"ahler
forms associated with corresponding the complex structures.
Then $\omega_J + \1 \omega_K$ is holomorphically symplectic with
respect to $I$.}
The Calabi-Yau theorem
\cite{_Yau_}
shows that, conversely, every compact holomorphically symplectic
K\"ahler manifold admits a hyperk\"ahler structure, which is uniquely
defined by these data.

\subsection[Quaternionic-Hermitian spaces and differential
forms over a hyperk\"ahler manifold.]%
{Quaternionic-Hermitian spaces and differential
forms \\over a hyperk\"ahler manifold.}

Let $M$ be a holomorphically symplectic K\"ahler manifold, and
\[ \c H = (I, J, K, (\cdot,\cdot)) \] be a hyperk\"ahler structure on 
$M$, which exists and is unique 
by the Calabi-Yau theorem. Consider the K\"ahler classes $\omega_I$, $\omega_J$,
$\omega_K\in H^2(M, \R)$ associated with $I$, $J$, 
$K$, and let $L_{\omega_\alpha}$, 
$\Lambda_{\omega_\alpha}:\; H^*(M) \arrow H^*(M)$,
$\alpha = I, J, K$, be the corresponding Hodge operators.

\hfill

\theorem \label{_so_5_Theorem_} 
(\cite{_so5_on_cohomo_}, Theorem 1) \ \ 
The operators $L_{\omega_\alpha}$, $\Lambda_{\omega_\alpha}$, 
$\alpha = I, J, K$, generate a 10-dimensional Lie
algebra $\g(\c H)$, which is naturally isomorphic to $\goth{so}(4,1)$.

\hfill

{\bf Sketch of a proof:} Consider the Hodge operators 
$L_{\omega_\alpha}, \Lambda_{\omega_\alpha}$, $\alpha = I, J, K$, 
acting on differential forms over $M$. These operators commute with
the Laplacian (by Kodaira) and hence act on cohomology. 
Therefore, to prove \ref{_so_5_Theorem_} it suffices to show that 
$L_{\omega_\alpha}$, $\Lambda_{\omega_\alpha}$ generate
a Lie algebra $\goth{so}(4,1)$ acting on differential forms. 

Let $T$ be a vector space equipped with a quaternionic action and
a positive definite $\R$-valued symmetric pairing $(\cdot,\cdot)$
such that the quaternions $I$, $J$, $K$ are orthogonal operators
with respect to $(\cdot,\cdot)$. Such a vector space is called
{\bf quaternionic-Hermitian}. For every quaternionic-Hermitian
vector space $T$, we can consider an action of the Lie algebra $\g(T)$
on $\Lambda^*_\R(T)$, defined in the same way as for differential
forms: the operators $L_{\omega_\alpha}$ act as exterior multiplication
by $\omega_\alpha$, and the operators $\Lambda_{\omega_\alpha}$
are adjoint to $L_{\omega_\alpha}$  with respect to the positive
definite metric induced by $(\cdot,\cdot)$.

Let $x\in M$, and $T_xM$ be the tangent space at $x$. 
Since $M$ is hyperk\"ahler, the space $T_xM$ is equipped with a natural
quaternionic-Hermitian structure. The action of $\g(\c H)$ 
on $\Lambda^*(M)$ comes from an action of $\g(\c H)$ 
on $\Lambda^*(T_x M)$. Therefore, to prove \ref{_so_5_Theorem_}
it suffices to show that for every quaternionic-Hermitian 
space $T$, the Lie algebra $\g(T)$ is isomorphic to
$\goth{so}(1,4)$.

 Let $T = \oplus T_i$ be an orthogonal decomposition
of $T$ onto a direct sum of $\Bbb H$-invariant subspaces.
Clearly, $\Lambda^*(T) = \bigotimes_i \Lambda^* T_i$,
and the action of $\g(\c H)$ on $\otimes \Lambda^* T_i$
is multiplicative with respect to this decomposition:

\begin{equation} \label{_g_multipli_Equation_}
\begin{split} \forall t \in \Lambda^*(T), 
   \;\; t &= 
   \otimes t_i, \; t_i \in \Lambda^*(T_i), \; \; h \in \g(\c H)\\[2mm]
   h(t) &= \sum_i t_1 \otimes ... \otimes h(t_i) \otimes ... t_n. 
\end{split}
\end{equation}
Therefore, it suffices to show that $\g(T)$ is canonically isomorphic
to $\goth{so}(1,4)$
for $\dim_{\Bbb H} T=1$. This is done by a computation.
\footnote{In \cite{_so5_on_cohomo_}, generators and relations of 
$\g(\c H)$ were written down, then a root system found. 
This root system turns out to be $B_2$, which gives an
isomorphism $\g(\c H)\otimes \C \cong \goth{so}(5, \C)$.
To find out what real form of $\goth{so}(5, \C)$
corresponds to $\g(\c H)$, we constructed a non-zero homomorphism 
$\g(\c H) \arrow \goth{sp}(1,1) \cong \goth{so}(1,4)$
(see also the proof of \ref{_center_Spin_action_Corollary_}).}
$\;\;\blacksquare$

\hfill

\corollary\label{_center_Spin_action_Corollary_} 
Consider the 
representation $\mbox{Spin}(4,1) \arrow End(H^*(M))$ corresponding to 
the action of $\g(\c H) \cong \goth{so}(4,1)$ on $H^*(M)$. 
Let $Z\cong \Z/ 2\Z$
be the center of $\mbox{Spin}(1,4)$, and $\iota$ be the non-trivial
element of $Z$. Then $\iota(\omega)= (-1)^{\dim\omega} \omega$.
In particular, $Z$ acts trivially on $H^{\even}(M)$.

\hfill

{\bf Proof:} Using \eqref{_g_multipli_Equation_}
as in the proof of \ref{_so_5_Theorem_}, we reduce
\ref{_center_Spin_action_Corollary_} to the case of action of
$\mbox{Spin}(1,4)$ on $\Lambda^*_\R(T)$, where $T$ is 
a quaternionic-Hermitian space, $\dim_{\Bbb H} T =1$.
Let $G(T)\subset \Lambda^*_\R(T)$ be the Lie group 
corresponding to $\g(T)$. Consider an action of $G(T)$ on the
space $W=\Lambda^{\odd}_\R(T)$ for $\dim_{\Bbb H}T =1$.
Let $(\cdot,\cdot)$
be the bilinear symmetric form on $W$ of signature $(4,4)$,
\[ 
   a, b {\arrow} (-1)^{\frac{\dim a+1}{2}} a\cup b. 
\]
Since $\Lambda^3_\R(T) \cong \Lambda^1_\R(T)^*$ and
$T$ is quaternionic, $W$ is equipped with a natural
quaternionic structure.
It is easy to check that $G(T)$ preserves $(\cdot,\cdot)$
and is generated by quaternionic matrices (see e. g. 
\cite{_so5_on_cohomo_}, Appendix).
This defines a non-trivial map $G(T) \arrow \mbox{Sp}(W)\cong \mbox{Sp}(1,1)$.
The  groups $G(T)$, $Sp(1,1)$ are simple and have the same
dimension. Therefore, the natural map $G(T) \arrow \mbox{Sp}(W)$
is an isomorphism.
This gives an explicit way to identify $G(T)$ and
$\mbox{Sp}(1,1)$. 

We obtained that the Lie algebra
$\g(T) \cong \goth{sp}(1,1)$ acts on 
$\Lambda^{\odd}_\R(\Bbb H)$ as on its fundamental 
representation. Since $\mbox{Sp}(1,1)\cong \mbox{Spin}(1,4)$,
and the central element of $\mbox{Sp}(1,1)$ acts on its fundamental
representation by $-1$, we obtain that
$\iota\restrict{\Lambda^{\odd}_\R(\Bbb H)}= -1$.
Similarly one checks that $\iota$ acts trivially
on $\Lambda^{\even}_\R(\Bbb H)$. $\;\;\blacksquare$

\subsection{Appendix. Generators and relations for
the Lie algebra $so(1,4)$.}

For the benefit of the reader, we give the relations
in $\g(T)$, computed  in \cite{_so5_on_cohomo_}.
We abbreviate $L_{\omega_I}$, $L_{\omega_J}$,
$L_{\omega_K}$,  $\Lambda_{\omega_I}$, $\Lambda_{\omega_J}$,
$\Lambda_{\omega_K}$ by $L_1$, $L_2$, $L_3$, $\Lambda_1$,
$\Lambda_2$, $\Lambda_3$. Let 
$K_{ij}:= [L_i,\Lambda_j]$, $i\neq j$, and $H:\; \Lambda^*(T) \arrow
\Lambda^*(T)$ act on $\Lambda^i(T)$ as  multiplication by
the scalar $\dim_\R T -2i$.
Then the following relations are true and define
the Lie algebra $\g(T)$ is a unique way:

\begin{equation} \label{_so5_relations_Equation_}
\begin{array}{l}
{}[ L_i, L_j] = [\Lambda_i,\Lambda_j] =0;\;\; \\[2mm]
[L_i,\Lambda_i]= H;\;\; [H, L_i] = 2 L_i;
\;\; [H, \Lambda_i]=-2\Lambda_i \\[2mm]{}
K_{ij}=-K_{ji}, [K_{ij}, K_{jk}]=2 K_{ik}, [K_{ij}, H] =0 \\[2mm]{}
[K_{ij} L_j]=2 L_i;\;\; [K_{ij} \Lambda_j] = 2 \Lambda_i \\[2mm]{}
[K_{ij}, L_k] = [K_{ij}, \Lambda_k] =0\;\; (k\neq i,j)\\[2mm]
\end{array}
\end{equation}


\section{Operator of Serre duality.}
\label{_operator_Serre_dua_Section_}


Let $M$ be a holomorphically symplectic manifold of K\"ahler
type, $\dim_\C M =n$. The canonical class of $M$ 
is naturally trivialized by the form $\Omega^{n/2}$, where
$\Omega$ is the holomorphic symplectic form.
Let $\Omega^i(M) \cong (\Omega^{n-i} M)^*$ be the duality coming
from this trivialization of the  canonical class,
and $\Omega^i(M) \cong (\Omega^i(M))^*$ be the duality defined
by the holomorphically symplectic form. Combining these two
maps, we obtain an isomorphism $\Omega^i(M) \cong \Omega^{n-i} M$.
The corresponding map $\eta^{i,j}:\; H^i(\Omega^j(M)) \arrow
H^{i}(\Omega^{n-j}(M))$
of cohomology is called  {\bf the Serre
duality homomorphism}.
Let $\eta = \bigoplus_{i,j} \eta^{i,j}$.
Clearly, $\eta$ is an invertible endomorphism of the space
$H^*(M)$. \footnote{The operator $\eta$ is an involution,
as follows from the argument one uses to show that
the standard Hodge operator $*$ is an involution.}
Yukawa multiplication $\bullet_{{}_Y}$ 
can be expressed via $\eta$ and 
the usual multiplication $\cup$ in cohomology as

\begin{equation} \label{_Yukawa_via_eta_Equation_} 
 h_1 \bullet_{{}_Y} h_2 = \eta^{-1}\bigg( \eta(h_1) \cup \eta(h_2)\bigg). 
\end{equation}

{\bf Remark:} The multiplication $\bullet_{{}_Y}$ is
determined by the complex structure and the section
of the canonical class, and the operator $\eta$ depends
on the choice of holomorphic symplectic form.

\hfill

Let $\c H = (I, J, K, (\cdot,\cdot))$ be a hyperk\"ahler structure on $M$,
and $G(\c H, \C) \subset End(H^*(M, \C))$ be the Lie group corresponding
to the Lie algebra $\g(\c H)\otimes \C = \goth{so}(1,4)$ of 
\ref{_so_5_Theorem_}.

\hfill

\theorem \label{_Serre_dua_through_g(H)_Theorem_} 
Consider the Serre duality operator $\eta$
acting on $H^*(M)$. Then $\eta$
belongs to $G(\c H, \C)$.

\hfill

The proof of \ref{_Serre_dua_through_g(H)_Theorem_}
takes the rest of this section.

\hfill

It is possible to describe explicitly the element of 
$G(\c H, \C)$ which corresponds to $\eta$. 
Let $T$ be a quaternionic-Hermitian space, and $\g(T) = \goth{so}(1,4)$
be the Lie algebra defined in the proof of \ref{_so_5_Theorem_}.
Let $\Lambda_\R^*(T)\otimes \C = \oplus \Lambda^{p,q}_I(T)$
be the Hodge decomposition of $\Lambda_\R^*(T)$ taken with respect
to the complex structure $I$, and $ad\,I$, $H$ be endomorphisms
of $\Lambda_\R^*(T)\otimes \C$, defined on 
$\Lambda^{p,q}_I(T)\subset \Lambda_\R^*(T)\otimes \C$ by 

\[ \omega^{p,q}\stackrel{ad\,I}{\arrow} (p-q) \1\omega^{pq}, \]

\[ \omega^{p,q}\stackrel{H}{\arrow} (p+q-2n) \omega^{pq}, \]
where $n= \dim_{\Bbb H} T$.

\hfill

\lemma \label{_adI,H_genera_Cartan_Lemma_} 
\cite{_so5_on_cohomo_}\ \  The endomorphisms
$ad\,I$, $H$ belong to $\g(T)$. These operators
generate a Cartan subalgebra\footnote{Which is 2-dimensional,
because $\g(T)\cong \goth{so}(1,4)$ with the root system $B_2$.}
 $\goth h$ for $\g(T)$. The root system of $\goth h$
is 

\begin{equation}\label{_root_sy_for_g(H)_Equation_} 
  \pm H, \ \ \pm \1 ad\,I, \ \   \pm H\pm \1 ad\,I. 
\end{equation}

$\blacksquare$

\hfill

In the above situation, let $w$ be the Weyl group element which
maps 

\begin{equation} \label{_Theta_as_Weyl_elt_Equation_}
  \begin{split} H - \1 ad\,I &\text{\ \  to \ \ } -H + \1 ad\,I\\ 
  \text{and} \ \ \ \ \ \ \ \ \ \ \ \ \ \ \ \ \  &\\
  H + \1 ad\,I &\text{\ \  to \ \ } H + \1 ad\,I.
  \end{split}
\end{equation}
This element defines
an automorphism $\Theta$ of the root system,
and therefore an automorphism of $\g(T)$.
Since Dynkin diagram of $\g(T)$ is $B_2$, 
all automorphisms of $\g(T)$ are inner, by well-known computation
of $Out(G)$ for classical groups (see, for instance, \cite{_Vin_Onishch_}).
Therefore, we may consider $\Theta$ as an element
of $G(T)/Z$, where $G(T)\subset End(T)$ is the Lie
group corresponding to $\g(T)$ and $Z$ is centre of $G(T)$.
Further on, we consider $\Theta$ as an element in $G(T)$
acting on $\g(T)$ as we just described. Since $Z = (\pm)$ 
(also \cite{_Vin_Onishch_}),
this involves a choice of one of two possible $\Theta\in G(T)$.
We specify this choice shortly thereafter.

\hfill

We describe the
isomorphism $\g(\c H) \cong \goth{so}(1,4)$ explicitly.
Let 
\begin{equation}\label{_V_definition_Equation_}
  V = V_0 \oplus V_2 \oplus V_4
\end{equation}
 be a 5-dimensional
graded vector space over $\R$, $\dim V_0 = \dim V_4 =1$,
$\dim V_2 =3$. Let $(\cdot,\cdot):\; V \times V \arrow \R$ 
be a bilinear symmetric form 
satisfying the following conditions

\hfill

(i) $(\cdot,\cdot)\restrict{V_2}$ is positiv definite.

(ii) $(V_0, V_2) = (V_4, V_2) = (V_0, V_0) = (V_4, V_4) =0$.

\hfill

Such form a exists and is unique up to a graded automorphism.
It has signature $(1,4)$. We define an isomorphism 
$i:\; \g (\c H) \arrow \goth{so}(V)$ as follows.

Let $V^o\subset H^2(M, \R)$ be the space spanned by $\omega_I$,
$\omega_J$, $\omega_K$. Let 
\[ 
   (\cdot,\cdot)_{\c H}:\; H^2(M) \times H^2(M) \arrow \R
\]
be the natural (Bogomolov--Beauville)
pairing considered in Section \ref{_g(M)_Section_}.
By definition (see \cite{_main_}), the form
$(\cdot,\cdot)_{\c H}$ is positive definite on $V^o$.
We identify $V_2$ with $V^o$. Let $\Bbb I$ be a generator
of $V_0$ and $\Upsilon$ be a generator of $V_4$,
such that $(\Bbb I, \Upsilon) =1$, $\alpha$ be an index 
which runs through $I$, $J$, $K$, 
and $\omega$ be an arbitrary vector in $V^o$. We define an action
of $L_{\omega_\alpha}$, $\Lambda_{\omega_\alpha}$
on $V$ as follows. 
\begin{equation}\label{_g(H)_action_on_V_Equation_}
\begin{array}{ll}
\mbox{(i)}&\ \  L_{\omega_\alpha} V_4 = \Lambda_{\omega_\alpha} V_0 =0\\[1mm]
\mbox{(ii)} &\ \ L_{\omega_\alpha} \omega = (\omega_\alpha, \omega) \Upsilon\\[1mm]
\mbox{(iii)} &\ \ \Lambda_{\omega_\alpha} \omega = 
 (\omega_\alpha, \omega) \Bbb I\\[1mm]
\mbox{(iv)} & \ \  L_{\omega_\alpha} \Bbb I = \omega_\alpha, \ 
     \Lambda_{\omega_\alpha}  \Upsilon  = \omega_\alpha.
\end{array}
\end{equation}

\hfill

An easy argument (see, e. g., \cite{_main_}, Theorem 8.1) implies that
 \eqref{_g(H)_action_on_V_Equation_} defines
an action of $\g(\c H)$ on $V$. \footnote{Let $T$ be a quaternionic-Hermitian
space, $\dim_{\Bbb H}T=1$. As a representation
of $\g(\c H)= \g(T)$, the space $V$ is canonically isomorphic to 
$\g(T) \Lambda^0(T) \subset \Lambda^*(T)$, where $\g(T) \Lambda^0(T)$
is a minimal $\g(T)$-invariant subspace of $\Lambda^*(T)$ containing
$0$-forms.}

\hfill

 Let $\Omega \in V^o\otimes \C$
be the form $\Omega = \omega_J + \1 \omega_K$, and
$\Theta_0:\; V\otimes \C \arrow V \otimes \C$ be the endomorphism
defined by

\[
   \Theta_0 (\Bbb I) = \Omega,\;\; \Theta_0(\bar \Omega) = \Upsilon,\;\;
   \Theta_0(\omega_I) = \omega_I, \; \;\Theta_0(\Omega) = \Bbb I,
   \Theta_0(\Upsilon) = \bar\Omega.
\]
Clearly, $\Theta_0$ is orthogonal and $\det\Theta_0 =1$.
Since $\left(G(\c H)\otimes \C\right)/Z= SO(5, \C)$, every
orthogonal automorphism $\gamma$ of $V\times \C$, $\det \gamma =1$
corresponds to an element of $G(\c H, \C)/Z$.
Therefore, we may consider $\Theta_0$ as an element of 
$G(\c H)/Z= Aut(G)= SO(1,4)$.

\hfill

\claim \label{_Theta_Theta_0_Claim_} 
The adjoint action of $\Theta_0$ on $\g(\c H)$ satisfies 
\eqref{_Theta_as_Weyl_elt_Equation_}.

{\bf Proof:} A calculation. $\;\;\blacksquare$

\hfill

It is also possible to check by calculation 
that $\eta$ acts on $H^*(M)$ normalizing
$\g(\c H)$ and the resulting automorphism of $\g(\c H)$ coincides
with $\Theta_0$ (see \ref{_eta_and_g(M)_Proposition_}). 
However, $\mbox{Aut}(\g(\c H))\cong \mbox{Sp}(1,1)/(\pm 1)$,
while $G(\c H) \cong \mbox{Sp}(1,1)$, unless there is no odd-dimensional
cohomology. To give an explicit statement of
\ref{_Serre_dua_through_g(H)_Theorem_}, we need to
get rid of the ambiguity in sign.

\hfill

Let $T$ be a quaternionic-Hermitian space, $\dim_{\Bbb H} T=1$,
and $W= \Lambda^\odd_\R(T)$ be the representation of
$\mbox{Sp}(1,1)$ defined in the proof of 
\ref{_center_Spin_action_Corollary_}. Let
$L_{\Omega}, \Lambda_\Omega:\; W\otimes \C \arrow W\otimes \C$,
$L_{\Omega} = L_{\omega_J} +\1 L_{\omega_K}$,
$\Lambda_{\Omega} = \Lambda_{\omega_J} +\1 \Lambda_{\omega_K}$.
Let $L_{\bar\Omega}, \Lambda_{\bar\Omega}:\; W\otimes \C \arrow W\otimes \C$
be the complex conjugate operators to $L_\Omega$, $\Lambda_\Omega$,
$L_{\bar\Omega} = L_{\omega_J} -\1 L_{\omega_K}$,
$\Lambda_{\bar\Omega} = \Lambda_{\omega_J} -\1 \Lambda_{\omega_K}$.
Let $\goth M_\Omega$, 
$\goth M_{\bar \Omega}:\; W\otimes \C \arrow W\otimes \C$
be the maps

\[ t \stackrel {\goth M_\Omega} \arrow (L_\Omega + \Lambda_{\bar \Omega}) t, \]

\[ t \stackrel {\goth M_{\bar\Omega}} \arrow (L_{\bar\Omega} + 
  \Lambda_{\Omega}) t.
\]

Let $W = W^a \oplus W^b$ be the decomposition defined by 
$W^a = \Lambda^{1,0}(T)\oplus \Lambda^{1,2}(T)$,
$W^b = \Lambda^{0,1}(T)\oplus \Lambda^{2,1}(T)$.
Since $\Omega= \omega_J +\1 \omega_K$ is a (2,0)-form, and $\bar\Omega$
is a (0,2)-form, 

\begin{itemize}
\item $L_{\Omega}$ maps $\Lambda^{p,q}(T)$ to $\Lambda^{p+2,q}(T)$
\item $\Lambda_{\bar \Omega}$ maps $\Lambda^{p,q}(T)$ to $\Lambda^{p-2,q}(T)$
\item $L_{\bar \Omega}$ maps $\Lambda^{p,q}(T)$ to $\Lambda^{p,q+2}(T)$
\item $\Lambda_{\Omega}$ maps $\Lambda^{p,q}(T)$ to $\Lambda^{p,q-2}(T)$.
\end{itemize}
Therefore, $L_{\Omega}$, $\Lambda_{\bar\Omega}$ vanish on $W^a$
and $L_{\bar\Omega}$, $\Lambda_{\Omega}$ vanish on $W^b$.
The triples $L_{\Omega}$, $\Lambda_{\bar\Omega}$, $H - \1 ad\,I$
and $L_{\bar\Omega}$, $\Lambda_{\Omega}$, $H + \1 ad\,I$
satisfy relations of $\goth{sl}(2)$ for the standard
vectors $e,f, h\in \goth{sl}(2)$. 
Clearly, $W^b$ is a representation of weight
one for the $\goth{sl}(2)$ generated by 
$L_{\Omega}$, $\Lambda_{\bar \Omega}$, $H - \1 ad\,I$,
and $W^a$ for $L_{\bar\Omega}$, $\Lambda_{\Omega}$, $H + \1 ad\,I$.
A computation of the action of $f+e$ on $\goth{sl}(2)$-representations
of weight one implies that $(\goth M_\Omega)^2\restrict{W^b} = Id$
and $(\goth M_{\bar \Omega})^2\restrict{W^a} = Id$. 
Let $\underline\Theta\in End (W\otimes \C)$,

\[ \underline\Theta (t) =\goth M_\Omega+ \goth M_{\bar \Omega}^2. \]

\lemma \label{_Theta_in_Sp_Lemma_}
The endomorphism $\underline\Theta\in End (W\otimes \C)$
belongs to $Sp(1,1)\otimes \C \subset End (W\otimes \C)$.

\hfill

{\bf Proof:} Define a scalar product $(\cdot,\cdot)$ on $\Lambda^{\odd}(T)$
as follows. Let $\epsilon:\; \Lambda^*(T) \arrow \R$ be the standard
projection to $\Lambda^4(T)\cong \R$, and 
\[ 
   \langle\cdot,\cdot\rangle:\;
   \Lambda^{\odd}(T) \times \Lambda^{\odd}(T) \arrow \R
\] 
be the Poincar\'e form defined by $\langle x,y\rangle = \epsilon(x\wedge y)$.
We define 
\begin{equation*}
\begin{split}
   (x,y) &= \langle x,y\rangle, \ \ \text{for} \ \ \dim x =1, \dim y =3, \\
      &= -\langle x,y\rangle, \ \ \text{for} \ \ \dim x=3, \dim y =1.
\end{split}
\end{equation*}
According to the standard construction of the $Sp(1,1)$-action
on $\Lambda^{\odd}(T)$ (see \cite{_so5_on_cohomo_}, Appendix, 
or any standard textbook on classical groups, like 
\cite{_Vin_Onishch_}), 
the group $Sp(1,1)$ is identified with the group of all
endomorphisms of $W$ which belong to $End_{\Bbb H}(W)$,
preserve the scalar product $(\cdot,\cdot)$ and have determinant 1.
From construction, it is clear that
$\underline\Theta\in End_{\Bbb H}(W)\otimes \C$.
The eigenvalues of $\underline\Theta$ are $-1$ of multiplicity
$2$ and $1$ of multiplicity $6$. Therefore, 
$\det \underline\Theta=1$. The decomposition 
$W= W^a \oplus W^b$ is orthogonal, and $\underline\Theta$
is the identity on $W^a$. Therefore,
to prove that $\underline\Theta\in G(T, \C)$ it remains to 
show that $\underline\Theta\restrict{W^b}$ preserves
the scalar product in $W^b$. Let $z_1, z_2$ be an orthonormal 
basis in $\Lambda^{1,0}(T)$. Then
$W^b$ is spanned by $e_1^1 := \bar z_1$, $e_1^2:=\bar z_2$,
$e_1^2:=\bar z_1 \wedge \Omega$, $e_2^2:= \bar z_2 \wedge \Omega$,
where $\Omega = z_1\wedge z_2$. Let $\goth{d}(i,j)$ equal 1
if either $i=1$, $j=2$ or $i=2$, $j=1$ and 0 otherwise.
Let $p:\; \{1,2\} \arrow \{1,2\}$ 
be the function defined by $p(2)=1$, $p(1)=2$.
As the definition implies, 

\[ 
   (e_i^j, e_k^l) = \goth{d}(i,k) \goth{d}(j,l). 
\]
and $\goth M_\Omega(e_i^j) = e_{p(i)}^j$.
This implies that $\underline\Theta\restrict{W^b}$
is orthogonal. \ref{_Theta_in_Sp_Lemma_} is proven.
$\;\;\blacksquare$

\hfill

Let 
\[ 
   i:\; G(T,\C) \arrow G(\c H, \C), \ \ G(T)= Sp(1,1)
\] 
be the natural 
homomorphism.\footnote{Which is an isomorphism unless $H^\odd(M)$
is empty.}
Denote the element $i(\underline\Theta)\in G(\c H, \C)$ 
by $\Theta$. The precise version of \ref{_Serre_dua_through_g(H)_Theorem_}
follows.

\hfill

{\bf \ref{_Serre_dua_through_g(H)_Theorem_}$'$}\   
Consider $\eta$, $\Theta$ as endomorphisms of $H^*(M, \C)$. Then

\begin{equation}\label{_eta_is_Theta_Equation_}
   \eta = \Theta. 
\end{equation}

\hfill

{\bf Proof:} Here is the plan of the proof.
As in the proof of \ref{_so_5_Theorem_},
we reduce \eqref{_eta_is_Theta_Equation_} to the similar
equation which holds for exterior forms over
a quaternionic-Hermitian space. The Serre duality operator can be 
defined on forms, and then it is shown that it carries
over to harmonic forms, which are the cohomology.
We can define the ``formal'' Serre duality operator 
\[ \eta^{p,q}:\; \Lambda^{p,q}_I(T)\arrow \Lambda^{2n-p,q}_I(T),\] 
where $T$ is a quaternionic-Hermitian space, $\dim_{\Bbb H} T =n$. 
The group $G(\c H, \C)$ acts on $\Lambda^*_\R(T)\otimes \C$.
Therefore, we can speak of action
of $\Theta$ on this space. Now, to prove 
\ref{_Serre_dua_through_g(H)_Theorem_}$'$ it will suffice to prove
\eqref{_eta_is_Theta_Equation_} for $\eta$, $\Theta$ acting
in $\Lambda^*_\R(T)\otimes \C$.

\hfill

Let $T$ be a quaternionic-Hermitian space,
and $\Lambda_\R^*(T) \otimes \C = \oplus \Lambda^{p,q}_I(T)$
be the Hodge decomposition corresponding to an operator
$I\in \Bbb H$ considered as a complex structure on $T$. We define
the Serre duality operator
\[ \eta^{p,q}:\; \Lambda^{p,q}_I(T)\arrow \Lambda^{2n-p,q}_I(T)\] 
as follows. Let $\Omega \in \Lambda^{2,0}(T)$,
$\Omega = \omega_J +\1 \omega_K$. Then $\Omega$
defines a non-degenerate $\C$-linear pairing
$\Lambda^{1,0}_I(T) \times \Lambda^{1,0}_I(T) \arrow \C$.
By multiplicativity, this defines a non-degenerate pairing

\begin{equation}\label{_i,0xi,0_pairing_Equation_}
  \Lambda^{i,0}_I(T) \times \Lambda^{i,0}_I(T) \arrow \C.
\end{equation}
Let $n = \dim_{\Bbb H}(T)$.
Then the $n$-th exterior power of $\Omega$ is a non-zero 
vector in the one-dimensional space $\Lambda^{2n,0}_I(T)$.
Using the form $\Omega^n$, we define a non-degenerate
pairing 

\begin{equation}\label{_ix2n-i_pairing_Equation_}
   \Lambda^{i,0}_I(T)\times \Lambda^{2n-i,0}_I(T) \arrow \C
\end{equation}
Considering \eqref{_i,0xi,0_pairing_Equation_},
\eqref{_ix2n-i_pairing_Equation_} as isomorphisms
between $\Lambda^{*,0}(T)$ and $\left(\Lambda^{*,0}(T)\right)^*$,
we can form the composition of \eqref{_i,0xi,0_pairing_Equation_}
and \eqref{_ix2n-i_pairing_Equation_}, which is a map

\[ 
  \eta^{i,0}_T:\; \Lambda^{i,0}(T) \arrow \Lambda^{2n-i,0}(T). 
\]
Using the canonical isomorphism 
$\Lambda^{p,q} \cong \Lambda^{p,0}\otimes \Lambda^{0,q}$,
we define

\[ \eta^{p,q}_T:\; \Lambda^{p,q}(T) \arrow \Lambda^{2n-p,q}(T),\;\;
   \eta^{p,q}_T(\omega^{p,0}\otimes \omega^{0,q}) :=
   \eta^{p,0}_T(\omega^{p,0})\otimes \omega^{0,q}.
\]
Let 

\begin{equation} \label{_eta_T_defini_Equation_}
   \eta_T := \oplus \eta^{p,q}_T:\; 
   \Lambda^*_\R(T)\otimes \C\arrow \Lambda^*_\R(T)\otimes \C.
\end{equation}

\hfill

\proposition \label{_eta_for_lin_spa_Proposition_} 
Let $M$ be a compact hyperk\"ahler manifold, of complex dimension $2n$.
Let $\eta:\; H^{p,q}_I(M) \arrow H^{2n-p,q}_I(M)$ be the Serre
duality operator associated with the complex structure $I$ and a
holomorphic symplectic form $\Omega = \omega_J + \1 \omega_K$.
Consider the tangent bundle of $M$ as a bundle of quaternionic-Hermitian
spaces, and let 
$\eta_T:\; \Lambda^*_\R(M)\otimes \C\arrow \Lambda^*_\R(M)\otimes \C$
be the operator defined fiberwise as in \eqref{_eta_T_defini_Equation_}.
Then $\eta_T$ preserves harmonic forms, and the resulting 
endomorphism of $H^*(M, \C)$ coincides with $\eta$.

{\bf Proof:} Follows from the definition. $\;\;\blacksquare$

\hfill

Return to the proof of \ref{_Serre_dua_through_g(H)_Theorem_}$'$.\ 
Let $T$ be a quaternionic-Hermitian space and $\eta_T$, $\Theta$
be the endomorphisms of $\Lambda^*_\R(T)\otimes \C$ constructed above.
By \ref{_eta_for_lin_spa_Proposition_}, we need to show that
$\eta_T=\Theta$. Let $T = \oplus T_i$ be a decomposition
of $T$ to a direct sum of quaternionic-Hermitian subspaces.
The operators $\eta_T$ and $\Theta$ are multiplicative with respect
to this decomposition (for $\eta_T$ it follows from the definition, 
and for $\Theta$ from \eqref{_g_multipli_Equation_}):

\begin{equation} 
\begin{split}
   \forall t \in \Lambda^*(T),t &= 
   \otimes t_i, \; t_i \in \Lambda^*(T_i), \\[3mm]
   \eta_T(t) &= \eta_T(t_1) \otimes \eta_T(t_2) 
                     \otimes ... \otimes \eta_T(t_n), \\[3mm]
   \Theta(t) &=  \Theta(t_1) \otimes \Theta(t_2) \otimes ...
                  \otimes \Theta(t_n) 
\end{split}
\end{equation}
Therefore, we may assume that $\dim_{\Bbb H}T =1$. 
On $\Lambda^\odd(T)$, we have written the action of $\Theta$
explicitly in coordinates (\ref{_Theta_in_Sp_Lemma_}).
It is clear that $\eta$ acts on $\Lambda^\odd(T)$
exactly in the same way. 

Let $G_o = SU(2)$ be the group of unitary quaternions acting
on $\Lambda^*(T)$ by multiplication, and
$\Lambda^-(T)$ be the space of all $G_o$-invariant
2-forms.
Consider $\Lambda^\even(T)$ as a representation of $\g (T)$. It is easy
to check that $\Lambda^\even(T)$ decomposes as
\[ \Lambda^\even(T) = 
   \Lambda^-(T) \oplus \bigg(\g(T) \cdot \Lambda^0(T)\bigg),
\]
where 
\[ \dim \Lambda^-(T) = 3, \ \ 
   \dim \bigg(\g(T) \cdot \Lambda^0(T)\bigg) =5.
\]
 On the space $\Lambda^-(T)$, the Lie algebra $\g(T)$ acts trivially, and
therefore, $\Theta$ acts as the identity. Since $\dim_{\Bbb H} T =1$,
$\eta$ acts as the identity on $(1,0)$-forms. Therefore,
$\eta$ acts as the identity on $\Lambda^-(T)$.
As a representation of $\g(T)$, the space 
$\g(T)\cdot\Lambda^0(T)\subset \Lambda^{\even}(T)$
is isomorphic to $V$ of \eqref{_V_definition_Equation_}.
We have written explicitly the action of $\Theta$ on $V$.
It is easy to theck that $\eta$ preserves $\g(T) \cdot \Lambda^0(T)$
and its action on $\g(T) \cdot \Lambda^0(T)$ coincides with 
$\Theta$. This proves \ref{_Serre_dua_through_g(H)_Theorem_}.
$\;\;\blacksquare$

\hfill

{\bf Appendix.} Here we give a summary of the interaction between
$\eta$ and the standard Hodge operators acting on the cohomology of $M$.

\hfill

Let $M$ be a compact K\"ahler manifold.
For a K\"ahler class $\omega\in H^{1,1}_I(M)$, consider the
Hodge endomorphism $\Lambda_{\omega}:\; H^{i}(M) \arrow H^{i-2}(M)$.
A well-known linear algebra argument\footnote{\cite{_Bou:Lie_}, 
VIII $\S$ 11; see also \cite{_main_}, Proposition 7.1.} implies that
the operator $\Lambda_{\omega}$ depends only on 
the cohomology class $\omega$, and is independent of  theRiemannian
or complex structure on $M$. Let $S\subset H^2(M, \R)$ be the 
set of all $\omega$ for which there exists a complex structure $I$
for which $\omega$ lies in K\"ahler cone.
If $M$ is holomorphically symplectic,
\cite{_main_}, Lemma 12.1 implies that for all $a, b$, for which
$a,b, a+b$ lie in $S$, we have

\[ {(a+b,a+b)_{\c H}}\Lambda_{a+b} =
   {(a,a)_{\c H}}\Lambda_a + {(b,b)_{\c H}}\Lambda_b.
\]
In \cite{_main_}, Lemma 5.6, we proved that $S$ is open in $H^2(M, \R)$.
From the argument used in the proof of Lemma 5.6, \cite{_main_},
it is clear that for every open set $U\subset H^2(M, \R)$, there
exists $x\in S$ such that for all $y\in U$, we have $x+y\in S$.
Therefore, there exist an (obviously, unique) linear map
\[ \check \Lambda:\; H^2(M) \arrow End(H^*(M)) \] such that for all
$a\in S$,  we have 
\[ 
   \check \Lambda(a) ={(a,a)_{\c H}}\Lambda_a.
\]

\hfill

\definition \label{_Lambda_a_Defintion_} 
Throughout this paper, we set $\Lambda_a$ to be 
${(a,a)_{\c H}}^{-1}\check \Lambda(a)\in End(H^*(M))$ 
for all $a \in H^2(M, \C)$,
${(a,a)_{\c H}} \neq 0$.

\hfill

\proposition \label{_eta_and_g(M)_Proposition_} 
Let $M$ be a holomorphically symplectic manifold of K\"ahler type,
$\Omega$ a holomorphic symplectic form, $\eta$ the operator of Serre 
duality, $\omega$ a class in $H^{1,1}(M)$. For each automorphism
$x\in End(H^*(M))$, denote by $\eta(x)$ the endomorphism 
$\eta x \eta^{-1}$. 
Then

\begin{description}
\item[(i)] $\eta$ is an involution.

\item[(ii)] $\eta(L_\omega) = [\Lambda_\Omega,  L_\omega]$.

\item[(iii)] $\eta(\Lambda_\omega) = [\Lambda_{\bar \Omega},  L_\omega]$.

\item[(iv)] $\eta(L_{\bar \Omega}) = \Lambda_\Omega$, 
$\eta(L_{\bar \Omega}) = L_{\Omega}$, 
$\eta(\Lambda_{\bar \Omega}) = \Lambda_{\Omega}$, 

\end{description}
where $L_a:\; H^i(M) \arrow H^{i+2}(M)$ is the operator of multiplication
by $a\in H^2(M)$, and $\Lambda_\Omega$, $\Lambda_{\bar \Omega}$
 are the operators defined above.

\hfill

{\bf Proof:} The part (i) follows from the definition (compare $\eta$
with the Hodge operator $*$, which is also involutive).
To prove (ii) -- (iii), notice that
the K\"ahler cone is Zariski dense in $H^{1,1}(M)$.
Therefore, we may assume that $\omega$ is a K\"ahler class.
Then, (ii), (iii) is a consequence of standard identities
in the algebra $\g(\c H)\cong \goth{so}(1,4)$ associated
with the hyperk\"ahler structure. Part (iv) is straightforward. 
$\;\;\blacksquare$


\section{Yukawa VFA as an equivariant VFA (proofs).}
\label{_equiv_proofs_Section_}


The chief tool used in our calculations is the following theorem.

\hfill

\theorem \label{_structure_alge_for_coho_hyperkahe_Theorem_} 
(\cite{_main_}, Theorem 11.1)
Let $M$ be a compact holomorphically symplectic manifold. Assume 
that $\dim H^{2,0}(M)=1$.
Let ${\Bbb V}$ be the linear space $H^2(M, \R)$ equipped with a 
natural scalar product $(\cdot,\cdot)_{\c H}$ 
of \cite{_Beauville_}, Remarques, p. 775, and \cite{_main_}, Theorem 5.1.
Let $\g(M)\subset End(H^*(M))$ be the Lie algebra 
generated by $L_\omega$, $\Lambda_\omega$, where $\omega$ runs through
K\"ahler classes corresponding to all complex structures on $M$.
Then $\g(M)$ is isomorphic to $\goth{so}({\Bbb V}\oplus \goth H)$, where $\goth H$
is a 2-dimensional real vector space with hyperbolic quadratic form.%
\footnote{For a holomorphically symplectic manifold with $h^{2,0}(M)=1$,
$h^1(M)=0$, the signature of $(\cdot,\cdot)_{\c H}$ is $(3,m-3)$, for 
$m = h^2(M)$ (see \cite{_main_}). In this case, the Lie algebra 
$\g(M)$ is isomorphic to $\goth{so}(4, m-2)$.}

$\blacksquare$

\hfill

Let $G(M)\subset End(H^*(M))$ be the Lie group corresponding
to $\g(M)$. 

\hfill

\lemma \label{_G(M)_on_Theta_Proposition_} %
Let $M$ be a hyperk\"ahler manifold equipped with a
hyperk\"ahler structure $(I, J, K, (\cdot,\cdot))$, 
and $\Theta\in G(\c H)\subset G(M)$
be the group element constructed in Section 
\ref{_operator_Serre_dua_Section_}.
Let 

\begin{equation} \label{_ortho_in_H^11_Equation_}
   \omega_1, \omega_2\in H^{1,1}_I(M), \ \ 
   (\omega_1,\omega_I)_{\c H}=(\omega_2,\omega_I)_{\c H}=
   (\omega_1,\omega_2)_{\c H}=0. 
   \ \ 
\end{equation}
 Consider 
$g = [ L_{\omega_1}, \Lambda_{\omega_2}]$ as an element of the Lie algebra
$\g_0(M)\subset \g(M)$. Then $\Theta$ stabilizes $g$: 

\[ Ad(\Theta) g = g. \]

{\bf Proof:} Let $E_g\subset G(M)\otimes \C$
be the one-dimensional subgroup corresponding to $g$.
We need to show that for all $\gamma \in E_g$,
$\gamma$ commutes with $\Theta$. By \cite{_main_},
Proposition 13.2, for all $\omega\in H^2(M)$,
we have

\[ [g, L_\omega] = (\omega,\omega_1)_{\c H} L_{\omega_2} - 
   (\omega,\omega_1)_{\c H} L_{\omega_1},
\]
and
\[  [g, \Lambda_{\omega}] = -(\omega,\omega_1)_{\c H} 
    \Lambda_{\omega_2} +
    (\omega,\omega_1)_{\c H} \Lambda_{\omega_1}.
\]
Since $H^{2,0}_I(M)$ is orthogonal to $H^{1,1}_I(M)$,
we obtain that 
\[ 
   [g,L_{\omega_\alpha}]=[g,\Lambda_{\omega_\alpha}] =0,
\]
for $\alpha = J, K$. Since $\omega_i$ are orthogonal to $\omega_I$,
we know that $[g,L_{\omega_I}]=[g,\Lambda_{\omega_I}] =0$.
Therefore, for $\gamma\in E_g$,
the automorphism 
\[ 
    Ad(\gamma):\; \g(M)\otimes \C\arrow\g(M)\otimes \C 
\]
acts trivially on $\g(\c H)\otimes \C\subset \g(M)\otimes \C$. 
Therefore, $\gamma$ commutes with $\Theta$. This proves 
\ref{_G(M)_on_Theta_Proposition_}. $\;\;\blacksquare$

\hfill

\proposition \label{_eta_depends_on_peri_Proposition_} 
Let $M$ be a holomorphically symplectic manifold, and $(I_1, \Omega_1)$, 
$(I_2, \Omega_2)$  holomorphic
symplectic structures. 
Let \[ \eta_1, \eta_2: \; H^*(M) \arrow H^*(M) \]
be the Serre duality operators corresponding to 
\[ (I_1, \Omega_1),\ \  (I_2, \Omega_2).\] 
Assume that the cohomology classes
$[\Omega_1]$, $[\Omega_2] \in H^2(M, \C)$ are equal.
Then $\eta_1=\eta_2$.

\hfill

{\bf Proof:} Let $\c H_1$, $\c H_2$ be the hyperk\"ahler
structures which induce $(I_1, \Omega_1)$,  $(I_2, \Omega_2)$
respectively. Let 
$\omega_\alpha^i\in H^2(M, \R)$, $\alpha = I, J, K$, $i=1,2$ 
be the standard cohomology 
classes $\omega_I, \omega_J, \omega_K$ for the hyperk\"ahler
structure $\c H_i$. Since $[\Omega_1]=[\Omega_2]$, we have 
$\omega_K^1=\omega_K^2$ and $\omega_J^1=\omega_J^2$.
If $\omega_I^1=\omega_I^2$, then the Lie algebras $\g(\c H_1)$,  
$\g(\c H_2)\subset End(H^*(M))$
coincide. Therefore, the groups $G(\c H_1)$, $G(\c H_2)$
also coincide. Expressing $\eta_i$ in terms of $G(\c H_i)$
as in  \eqref{_eta_is_Theta_Equation_}, we obtain that
$\eta_1=\eta_2$. 

Using \ref{_g(M)=so_Theorem_}, 
it is easy to show that the Hodge decompositions
$H^{p,q}_I(M):=H^{p,q}_{I_1}(M)$, $H^{p,q}_{I_2}(M)$ coincide.
Let $K_i$, $i=1,2$ be the K\"ahler cones of the
complex structures $I_i$. If $K_1$ intersects $K_2$, we
may apply Calabi-Yau to find the hyperk\"ahler
structures $\c H_1$, $\c H_2$ satisfying 
$\omega_I^1=\omega_I^2$, such that $\c H_i$ induces
$(I_i, \Omega_i), i=1,2$. Applying the previous argument,
we find that in this case $\eta_1=\eta_2$.
However, it is {\it a priori} possible that
$K_1\cap K_2 =\emptyset$. In this case, we apply
the following argument.

\hfill

Let $\omega_1, \omega_2, \omega_3$ be classes in $H^2(M, \R)$
satisfying 
\begin{equation}\label{_omega_orthonormal_Equation_}
(\omega_1, \omega_1)_{\c H} = (\omega_2, \omega_2)_{\c H} =
(\omega_3, \omega_3)_{\c H}>0, \; (\omega_i, \omega_j)_{\c H}=0.
\end{equation}
Let $P\subset H^2(M, \R)$ be the linear span of $\omega_i$, $i=1,2,3$.
Let $\g(P)\subset \g(M)$ be the Lie algebra generated by $L_\omega$,
$\Lambda_\omega$, for $\omega\in P$. The argument proving 
\ref{_structure_alge_for_coho_hyperkahe_Theorem_} easily implies that
$\g(P)$ is isomorphic to $\goth{so}(1,4)$ in the same way
as $\g(\c H)$ is. Applying the standard construction, we
obtain a Serre duality operator 
$\Theta_{\omega_1, \omega_2, \omega_3}\in End(H^*(M))$
which is defined in the same way as $\Theta$ for $\g(\c H)$.

\hfill

\lemma\label{_Theta_inde_algebraic_Lemma_} 
Under assumptions of \ref{_eta_depends_on_peri_Proposition_},
let $S$ be the set of all cohomology classses 
$\omega\in H^{1,1}_I(M)$  satisfying
\begin{equation} \label{_ortho_for_omega_Equation_}
  (\omega, \omega)_{\c H} = (\omega_J, \omega_J)_{\c H}, \; 
  (\omega, \omega_J)_{\c H}=(\omega, \omega_K)_{\c H}=0, i\neq j.
\end{equation} 
Let $\omega,\omega' \in S$.
Then $\Theta_{\omega,\omega_J, \omega_K}= \Theta_{\omega',\omega_J, \omega_K}.$

\hfill

{\bf Proof:} For $\omega, \omega'\in K_1$, this is implied by
$\Theta_{\omega, \omega_J, \omega_K} = \eta_1$,
$\Theta_{\omega', \omega_J, \omega_K} = \eta_1$, which 
is a consequence of \ref{_Serre_dua_through_g(H)_Theorem_}$'$.
(see the argument in the beginning of 
\ref{_eta_depends_on_peri_Proposition_}, Proof)
Clearly, the set $S$ is connected, and $S\cap K_1$ is Zariski dense in $S$.
Since $\Theta_{\omega, \omega_J, \omega_K}$ is expressed
algebraically as a function of $\omega$, and is constant in 
a Zariski dense set, the map 
\[ \Theta_{\;\cdot\;, \omega_J, \omega_K}:\; S \arrow End(H^*(M)) \]
is constant. This proves 
\ref{_Theta_inde_algebraic_Lemma_} and
\ref{_eta_depends_on_peri_Proposition_} $\;\;\blacksquare$

\hfill

To prove \ref{_stabi_I_commu_with_Yuka_Theorem_},
\ref{_Yu_equiv_Theorem_}, we use the following claim:

\hfill

\claim\label{_eta_for_diff_c_H_equiv_Claim_} 
Let $\c H_1$, $\c H_2$ be hyperk\"ahler structures on $M$,
and $\eta_1$, $\eta_2$ be the corresponding Serre duality
operators.
Denote by $\omega_\alpha^i$, $\alpha = I, J, K$, $i=1,2$
the natural K\"ahler forms associated with $\c H_1$, $\c H_2$.
Let $g\in G_0(M)$ be an element such that under the natural
action of $G_0(M)$ on $H^2(M)$, $g(\omega_\alpha^1)=\omega_\alpha^2$.
Then $g \eta_1 g^{-1} = \eta_2$.

{\bf Proof:} Follows from \eqref{_eta_is_Theta_Equation_}.
$\;\;\blacksquare$

\hfill

Now, \ref{_stabi_I_commu_with_Yuka_Theorem_} is equivalent 
to the following statement:

\hfill

{\bf \ref{_stabi_I_commu_with_Yuka_Theorem_}$'$:} \ 
Let $M$ be a holomorphically symplectic manifold, $\dim_\C M = n$,
$[\Omega] \in H^2(M, \C)$ be the cohomology class of its holomorphic
symplectic form, $g\in G_0(M)$ be an 
element preserving the line passing through
$\Omega$.\footnote{By definition of $G_0^I(M)$, 
the group element $g$ preserves the line
passing through $[\Omega]$ if and only if $g \in G_0^I(M)$.}
Using \ref{_Theta_inde_algebraic_Lemma_} which maps $[\Omega]$
to $c[\Omega]$, $c \in \C$, we may canonically associate
the Yukawa product map on $H^*(M)$ with a cohomology class 
$\Omega_0\in H^2(M, \C)$, given that $\Omega_0$ is the class of a 
holomorphic symplectic form for some complex structure on $M$.
Let $Y_1, Y_2:\; H^*(M) \times H^*(M) \arrow H^*(M)$ be the Yukawa
products associated with the cohomology classes $[\Omega]$, 
$c [\Omega] \in H^{2n, 0}_I(M)$.
Then

\[ g Y_1 g^{-1}= Y_2. \]

{\bf Proof:} Let $\eta_{\Omega}$, $\eta_{\Omega'}$ be the 
operators $\eta$ associated with $\Omega$, $\Omega'$, where
$\Omega' = c \Omega$. \ref{_eta_for_diff_c_H_equiv_Claim_}
implies that $\eta_{\Omega'} = g \eta_\Omega g^{-1}$.
Now,

\[ Y_1(x,y) = 
   \eta_\Omega^{-1}\left(\eta_\Omega(x)\cup \eta_\Omega(y)\right) 
\]
and
\[ Y_2(x,y) = 
   \eta_{\Omega'}^{-1}\left(\eta_{\Omega'}(x)\cup \eta_{\Omega'}(y)\right). 
\]
Since $g\in G_0(M)$, the group element $g$ acts by automorphisms
on the (usual) cohomology ring. Therefore, 
\begin{equation*}
\begin{split}
   Y_2(x,y) &=   
   g \eta_\Omega^{-1} g^{-1}
   \left(g \eta_\Omega g^{-1}(x)\cup g \eta_\Omega g^{-1}(y)\right)\\ 
   &=  g \eta_\Omega^{-1}
   \left(\eta_\Omega g^{-1}(x)\cup \eta_\Omega g^{-1}(y)\right)\\ 
   & = g Y_1 (g^{-1} x, g^{-1}y).
\end{split}
\end{equation*}

\ref{_stabi_I_commu_with_Yuka_Theorem_}$'$ is proven. $\;\;\blacksquare$

\hfill

{\bf Proof of \ref{_Yu_equiv_Theorem_}:}
Let 
\[ \bullet_{I_1},\  \bullet_{I_2}:\; H^*(M) \times 
   H^*(M) \arrow H^*(M)\otimes H^{n,0}(M)
\] 
be the Yukawa multiplication
maps constructed by $I_1$, $I_2 \in Comp$, 
\[ H^{n,0}(M)= \underline K\restrict{I_1}=\underline K\restrict{I_2}.\]
\ref{_eta_depends_on_peri_Proposition_} implies that,
whenever $P_o(I_1) = P_o(I_2)$, we have $\bullet_{I_1}=\bullet_{I_2}$.
Therefore, there exist a tensor 
$\underline\bullet_{{}_Y}:\; B\times B \arrow B\otimes K$
such that $\bullet_{{}_Y}$ is obtained as a pullback of $Y$.

To prove \ref{_Yu_equiv_Theorem_}, it remains to show that
$\underline\bullet_{{}_Y}$ is equivariant with respect to the
natural $G_0(M)$-equivariant structure on $B$.
This is implied by \ref{_eta_for_diff_c_H_equiv_Claim_}.
$\;\;\blacksquare$

\nopagebreak
\section[Periods of hyperk\"ahler manifolds 
and Tian -- Todorov coordinates.]{Periods of hyperk\"ahler manifolds \\
and Tian -- Todorov coordinates.}\nopagebreak
\label{_Periods_and_Tia_Todo_coordi_Section_}
\nopagebreak


Let $M$ be a holomorphically symplectic manifold, and $Comp$
be its (marked, coarse) deformation space. We give the
following description of Tian--\-To\-do\-rov coordinates on $Comp$.

Let $\g_0(M)\subset End(H^*(M))$ be the Lie algebra
of \ref{_g(M)=so_Theorem_}, 
\[ \g_0(M) \cong \goth{so}\!\left(H^2(M,\R), \;
   (\cdot,\cdot)_{\c H}\right).\]
Let $I\in Comp$, and $ad\,I\in \g_0(M)$ be the corersponding endomorphism
of cohomology. Let $G_0(M)\subset End(H^*(M))$ be the Lie group
corresponding to $\g_0(M)$ and $X \subset \g_0(M)$, 
$X = G_0(M) \cdot ad\,I$ be the adjoint orbit containing 
$ad\,I$. As we have seen previously, the map 
\[ P_o:\; Comp \arrow X, \ \ J \arrow ad J\] is \'etale.

Let
\begin{equation} \label{_g_decompo_under_ad_I_Equation_}
   \g_0(M) \otimes \C = 
   \g_0^{I,-2}(M) \oplus\g_0^{I,0}(M) \oplus \g_0^{I,2}(M)
\end{equation}
be the decomposition of $\g_0(M) \otimes \C$ defined by 
$[ ad\,I, g] = \1 k g$, for all $g\in \g_0^{I,k}(M)$,
$k = -2, 0, 2$.  
Earlier, we used the notation $\g_0^{I}(M)$ for 
$\g_0^{I,0}(M) \cap \g_0(M)$. 

\hfill

\claim \label{_g^20=g^02=H^1,1_Claim_} 
There is a natural isomorphism of linear spaces

\[ 
   \g_0^{I,-2}(M) \cong  \g_0^{I,2}(M) \cong H^{1,1}(M), 
\]
provided by the maps 
\[ \alpha \arrow [ L_\alpha \Lambda_\Omega] \in \g_0^{I,-2}(M),\ \ 
   \alpha \arrow [ L_\alpha \Lambda_{\bar \Omega}] \in \g_0^{I,2}(M),
\]
for all $\alpha \in  H^{1,1}(M)$.

\hfill

{\bf Proof:} Let $\g(M)$ be the Lie algebra of 
\ref{_structure_alge_for_coho_hyperkahe_Theorem_}. The space $H^*(M)$
is graded by the dimensions of cocycles. Consider the associated
grading $\g(M) = \g_{-2}(M) \oplus \g_0(M) \oplus \g_2(M)$.
In \cite{_main_}, we proved that $\g_2(M)$ is the linear span
of all operators $L_\omega$, for all $\omega \in H^2(M, \R)$,
and $\g_{-2}(M)$ is the linear span of all $\Lambda_\omega$,
$\omega \in H^2(M, \R)$ for all $\omega$ such that $\Lambda_\omega$
is defined. Let $\g_i^{I, j}(M) \subset \g_i(M)\otimes \C$ be the space
of all $x \in \g_i(M)$ satisfying $[ ad\,I, x] = j \1$.
Clearly, $\g_2^{I, 0}(M)\otimes \C$ is the linear span of all 
operators $L_\omega:\; H^*(M) \arrow H^*(M)$, 
$\omega \in H^{1,1}(M)$. Similarly,
$\g_{-2}^{I, 0}(M)\otimes \C$ is the linear span of all $\Lambda_\omega$,
$\omega \in H^{1,1}(M)$, such that $\Lambda_\omega$
is defined. This identifies $\g_{\pm2}^{I, 0}(M)$ with 
$H^{1,1}(M)$.
As we show in Section \ref{_operator_Serre_dua_Section_},
the adjoint action of $\eta$ interchanges $\1 H$ and $ad\,I$,
where $H$ is the standard Hodge operator,
$H = [L_\omega, \Lambda_\omega]$. Therefore,
$\eta$ identifies $\g_i^{I, j}(M)$ with $\g_j^{I, i}(M)$.
Applying this to $\g_{\pm2}^{I, 0}(M) \cong H^{1,1}(M)$,
we obtain an identification of 
$\g_{0}^{I, \pm 2}(M)$ with $H^{1,1}(M)$.
Unravelling this definition and applying 
\ref{_eta_and_g(M)_Proposition_}, we obtain
\ref{_g^20=g^02=H^1,1_Claim_}.
$\;\;\blacksquare$

\hfill

Let $R:\; X \arrow {\Bbb P}(H^2(M, \C))$ be the map associating a line

\[ H^{2,0}_I(M) = \left\{ 
   \lambda \in H^2(M, \C) \ \ | \ \ ad\,I(x) = 2\1 x 
                 \right\}
\]
to $ad\,I \in X\subset G_0(M)$.

\hfill

\lemma \label{_P_c_isomo_Lemma_} 
The map $R:\; X \arrow {\Bbb P}(H^2(M, \C))$ is a closed, 
complex analytic embedding. The image of $R$ coincides with the
conic  $C\subset {\Bbb P}(H^2(M, \C))$,

\[ C = \left\{ l\in {\Bbb P}(H^2(M, \C)) \ \ | \ \ (l,l)_{\c H} =0 
   \right\}
\]
where $(\cdot,\cdot)_{\c H}$ is the canonical scalar 
product on $H^2(M)$ (Section \ref{_g(M)_Section_}).

{\bf Proof:} Follows from a trivial linear-algebraic argument
(\cite{_Todorov_}; see also \cite{_main_}, Section 6).
$\;\;\blacksquare$

\hfill

Let $G_0(M) \subset End(H^*(M))$ be the Lie group associated with
$\g_0(M)$. Consider the action of a group 
$G_0(M)\subset End(H^*(M))$ corresponding to
$\g_0(M) = \goth{so}\bigg(H^2(M); (\cdot,\cdot)_{\c H}\bigg)$
on $H^2(M)$. This defines an action of $G_0(M)\otimes \C$ on
$C \subset {\Bbb P}(H^2(M, \C))$.

\hfill

Consider the action of $\g^{-2, I}(M)\subset \g_0(M) \otimes \C$ on
$C = X$. The Lie algebra $\g^{-2, I}(M)$ is commutative. Thus, we obtain
a number of commuting holomorphic vector fields on $X$. 
Since the action of $G_0(M)$ on $X$ is transitive,
$\dim X = \dim \g^{-2, I}(M) = \dim H^{1,1}(M)$
and $G^{2,I}_0(M)$, $G^{0,I}_0(M)$ stabilizes $R(ad\,I)$, the 
action of the abelian Lie group $G^{-2, I}(M)$ defines coordinates in
a neighbourhood $U$ of $R(ad\,I)\in X$. We have obtained a natural open
embedding $i:\; B \hookrightarrow C =X$, where $B$ is an open
ball in $\g_0^{-2,I}(M)$. 

\hfill

\theorem \label{_Tian_Todo_through_peri_Theorem_}
Let $M$ be a holomorphically symplectic manifold, $X$ its moduli space and
 $i:\; B \hookrightarrow X$ be the open embedding constructed above.
Lift $i$ to a map $\tilde i:\; B \hookrightarrow Comp$. Identifying 
$\g_0^{2,I}(M)$ with $H^{1,1}(M)$ as in \ref{_g^20=g^02=H^1,1_Claim_},
we can realize $B$ as an open ball in $H^{1,1}(M)$. Then, for $B$
sufficiently small, $\tilde i$ coincides with the Tian--Todorov map
of \ref{_Bogo_Tian_Todo_Theorem_}.

\hfill

{\bf Proof of \ref{_Tian_Todo_through_peri_Theorem_}:} 
Let $\Omega\in H^{2,0}(M, \C)$ be the holomorphic symplectic
form of $M$, $v\in B \subset  \g_0^{-2,I}(M)$ and 
$e^v(\Omega)\in C$ be the point of $C$ corresponding to $v$.
Let $P_c:\; Comp \arrow C$ be the map associating 
the line $H^{2,0}_L(M)\in {\Bbb P}(H^2(M, \C))$ to the
complex structure $L\in Comp$. Let 
$\bar \Omega\in \Lambda^{0,2}(M)$ be the complex conjugate
to $\Omega$. The standard identification of $H^{1,1}(M)$
with $\g_0^{-2,I}$ goes as 

\[ \lambda 
   \stackrel \tau \arrow 
   [ L_\lambda, \Lambda_{\bar\Omega}], 
\]
where $\lambda\in H^{1,1}(M)$ and 
$L_\lambda$, $\Lambda_{\bar\Omega}\in End(H^*(M))$
are the standard Hodge operators of \cite{_main_}%
\footnote{The operator $L_\lambda$ multiplies $\eta \in H^*(M)$ by $\lambda$.
The operator $\Lambda_{\Omega}$ is equal to 
$\Lambda_{\omega_J} +\1 \Lambda_{\omega_K} $, where 
$\Lambda_{\omega_J}$, $\Lambda_{\omega_K}$ are the Hodge operators
of complex structures $J$, $K$, in a hyperk\"ahler structure 
$(I, J, K, (\cdot,\cdot))$
on $M$. See also \ref{_Lambda_a_Defintion_}.}
 associated with $\lambda$, $\bar\Omega$. Then
\ref{_Tian_Todo_through_peri_Theorem_} is equivalent to the following
explicit statement:
\begin{equation} \label{_Tian_Todo_through_peri_Equation_}
   \left( e^v(\Omega)\right)^{\Bbb P} = P_c( \phi(\tau^{-1}(v))), 
\end{equation}
where $\phi:\; B \arrow Comp$ is the Tian--Todorov map,
and 
\[ (\ \ )^{\Bbb P}:\; H^2(M, \C)\backslash 0 
   \arrow {\Bbb P}(H^2(M, \C)) 
\]
is the map associating to a vector 
$x\in H^2(M, \C)\backslash 0$ the line passing through $x$.

\hfill

The K\"ahler cone $K$ is open in $H^{1,1}(M) \cap H^2(M, \R)$,
and $H^{1,1}(M) \cap H^2(M, \R)$ is the set of fixed points of an
anti-complex involution of the affine space
$H^{1,1}(M)$. Therefore, every complex analytic map
on the open ball $B$ is determined by its values on $K\cap B$. 
Therefore, in proving \eqref{_Tian_Todo_through_peri_Equation_}, we may assume
that the cohomology class $\omega = \tau^{-1}(v)$ is a K\"ahler class
for some K\"ahler structure on $M$.

Let $\c H= (I, J, K, (\cdot,\cdot))$ be the 
hyperk\"ahler structure on $M$, 
inducing $I$, $\Omega$ and $\omega$:

\[ \omega =\omega_I, \ \ \Omega = \omega_J + \1 \omega_K.\] 

\hfill

\lemma \label{_a_n_for_hyperkaahler_Lemma_} 
Let $M$ be a hyperk\"ahler manifold and 
$\omega = \omega_I\in \Lambda^{1,1}(M)$
be its K\"ahler form. Let $a= \tilde \omega$ be the harmonic
section of $\Lambda^{0,1}(TM)$ corresponding to $\omega$.
Let $\phi(t\omega) = \sum a_i$,
$a_n = G_{\bar \6} \6 \sum_{i+j = n-1} a_i \bullet_{{}_Y} a_j$
be the image of $t a$ under the Tian-Todorov map,
$t\in \R$. Then $a_n =0$ for all $n>0$.

\hfill

{\bf Proof:} Clearly, it suffices to show that $a_1 =0$, where $a_1 =
G_{\bar \6} \6 \left(\tilde \omega \bullet_{{}_Y} \tilde \omega\right)$.
Consider the standard Serre duality operator 
$\eta:\; \Omega^{p,q}(M)\arrow\Omega^{n-p,q}(M)$. By definition,
$\tilde\omega= \eta(\omega)$, and 
$x \bullet_{{}_Y} y= \eta\left( \eta^{-1}(x) \cup  \eta^{-1}(y)\right)$.
Therefore, 

\[ a_1 =G_{\bar \6} \6 \tilde \omega \bullet_{{}_Y} \tilde \omega = 
   G_{\bar \6} \6 \eta(\omega \cup \omega).
\]
Since $\omega$ is a K\"ahler form, $\omega \cup \omega$ is harmonic
(Kodaira). The map $\eta$ is compatible with the Hermitian structure,
and commutes with $\bar \6$. Therefore, $\eta$ maps harmonic
forms to harmonic forms.
We obtain that $\eta(\omega \cup \omega)$ is also harmonic, 
and $\6 \eta(\omega \cup \omega) =0$. $\;\;\blacksquare$

\hfill

Let $A_{t\omega}$ be $t\omega\widetilde{\;\;\;}$ 
considered as a differential operator \eqref{_A_as_diff_o_Equation_}, where 
\[ t\omega\widetilde{\;\;\;}\in\Lambda^{0,1}(M, TM)\] is the $TM$-valued 
(0,1)-form associated with the $(1,1)$-differential form
\[ t\omega\in \Lambda^{0,1}(\Omega^1(M))\]
via the identification $TM \cong \Omega^1M$ provided by the 
holomorphic symplectic structure.
We obtain that, for $a$ as in \ref{_a_n_for_hyperkaahler_Lemma_}
and $\6_{new}$ as in \eqref{_A_as_diff_o_Equation_}
the operator $\6_{new}$ is equal to $\bar\6 + A_{t\omega}$.
 The following proposition computes this differential
operator in terms of $\6$ and the quaternion action.

\hfill

\proposition \label{_omega_I_as_diff_ope_Proposition_} 
Let $M$ be a manifold equipped with a hyperk\"ahler structure
$(I, J, K, (\cdot,\cdot))$. Consider $M$ as a K\"ahler manifold
with K\"ahler structure associated with $I$ and $(\cdot,\cdot)$.
Let $\omega= \omega_I$ be its K\"ahler form. Consider $\omega$
as a section of $\Lambda^{0,1}(\Omega^1(M))$. Let $\underline A$ be
the section of $\Lambda^{0,1}(M, TM)$ obtained from $\omega$ 
through the isomorphism
$TM \cong \Omega^1(M)$ provided by the holomorphically symplectic 
structure, and
$A:\; C^{\infty}(M) \arrow \Lambda^{0,1}(M)$ be a differential
operator corresponding to $\underline A$ as in 
\eqref{_A_as_diff_o_Equation_}. Since $J$ anticommutes with
$I$, the operator $J:\; \Lambda^1_\R(M)\arrow \Lambda^1_\R(M)$  maps
$(1,0)$-forms to $(0,1)$-forms. Therefore,
we may consider a composition $J \circ \6$ as a 
differential operator from functions to $(0,1)$-forms.
Then 
\begin{equation} \label{_A_from_Kaehle_as_diff_ope_Equation_}
  A = J \circ \6.
\end{equation}

{\bf Proof:} Since $A$ and $J \circ \6$ are operators of the first order,
and a hyperk\"ahler manifold is flat up to $O(r^2)$, it suffices to check
\eqref{_A_from_Kaehle_as_diff_ope_Equation_} when $M$ is a flat manifold.
Assume $M$ is flat.
Let $x_i, y_i$ be holomorphic coordinates on $M$, such that the forms
$dx_i, dy_i, \overline{d x_i}, \overline{d y_i}$ constitute an
 orthonormal basis in the bundle of $1$-forms, and the holomorphic 
symplectic form is written as
$\Omega = \sum_i d x_i \wedge d y_i$. Then 
$\omega= \sum dx_i\wedge  \overline{dx_i} +dy_i\wedge  \overline{dy_i}$,
the operator $A$ is written in coordinates
as 
\[ A(f) = \sum_i \frac{\6f }{\6 x_i} \overline{d y_i }-
   \sum \frac{\6f }{\6 y_i} \overline{d x_i } 
\]
and $\6$ as
\[ \6 f = \sum_i \frac{\6f }{\6 x_i} dx_i 
   + \sum_i \frac{\6f }{\6 y_i} dy_i \]
On the other hand, $J(d x_i) = \overline{d y_i }$,
and $J(dy_i) = - \overline{d x_i }$. 
This proves \ref{_omega_I_as_diff_ope_Proposition_}.
$\;\;\blacksquare$

\hfill

We obtain the following explicit description of the Tian--Todorov map.
Let $\omega \in B \subset H^{1,1}(M)$ be a form such that
$\omega = t\omega_I$
for a hyperk\"ahler structure $(I, J, K, (\cdot,\cdot))$. Then
the complex structure $\phi(\omega)$ is defined by the differential operator
\begin{equation} \label{_Tian_todo_ope_expli_Equation_}
   \bar\6_{new} = \bar\6 + t I \circ \6.
\end{equation}

After appropriate substitutions, (we substitute
$\phi(\alpha)$ for $\6+ t I \circ \6$ and $v$ for $t\omega_I$)
the equation \eqref{_Tian_Todo_through_peri_Equation_}
takes the form

\begin{equation} \label{_Tian_Todo_through_peri_for_hype_Equation_}
 \left(e^{t[L_{\omega_I}, \Lambda_{\Omega}]}(\Omega)\right)^{\Bbb P} =
 P_c(\bar\6 + t I \circ \6),
\end{equation}
where by $P_c(\bar\6 + t I \circ \6)$ we understand the
line $H^{2,0}_L(M)\subset H^2(M, \C)$ associated with a complex
structure $L$ defined by the operator 
$\bar\6_{new} = \bar\6 + t I \circ \6$. 
To prove \ref{_Tian_Todo_through_peri_Theorem_}, it remains to show that 
\eqref{_Tian_Todo_through_peri_for_hype_Equation_}
is true for all hyperk\"ahler structures.

\hfill

A calculation using identities \eqref{_so5_relations_Equation_}
shows that 
$[L_{\omega_I}, \Lambda_{\Omega}] \Omega = 2 \omega$, and
$[L_{\omega_I}, \Lambda_{\Omega}] \omega = \bar \Omega$.
Therefore, 

\begin{equation} \label{_exp_of_Omega_Equation_}
e^{t[L_{\omega_I}, \Lambda_{\Omega}]}(\Omega) =
 \Omega + 2t \omega + t^2 \bar \Omega
\end{equation}

\hfill

The algebra ${{\Bbb H}\otimes_\R\C} = Mat_\C(2)$ naturally acts on
$\Lambda^1(M)$. We extend this action to an action of the group
 ${{\Bbb H}\otimes_\R\C}^*\cong GL(2, \C)$ on $\Lambda^*(M)$.

\hfill

\claim \label{_HxC_action_on_D_Remark_}
For each $\alpha \in {{\Bbb H}\otimes_\R\C}^*$,
the operator 

\[ \alpha (\bar \6 + A_{t \omega}):\;\; 
   f\in C^\infty(M) \arrow \alpha(\bar \6 + A_{t \omega}) \in \Lambda^1(M) 
\]
induces the same holomorphic structure on $M$ as 
$\bar \6 + A_{t \omega}$. 

\hfill

{\bf Proof:} Let $\goth C$ be the kernel of $\bar \6 + A:\; C^\infty(M)
\arrow \Lambda^1(M)$. By definition, the function
$f\in C^{\infty}(M)$ is holomorphic with respect
to the complex structure defined by $\bar \6 + A$ if and only if $f$
lies in $\goth C$. On the other hand, 
$\ker \left(\alpha (\bar \6 + A_{t \omega})\right)$ coincides with
$\ker \left(\bar \6 + A_{t \omega}\right)$. $\;\;\blacksquare$

\hfill

The following claim gives a
description of the complex structure associated with
$\bar\6_{new} = \bar\6 + t I \circ \6$, in terms of the hyperk\"ahler
structure.

\hfill

\claim \label{_6_new_induced_by_hype_expli_Claim_} 
Let $M$ be a hyperk\"ahler manifold and 
\[ \bar\6_{new}= \bar \6 + t J \circ \6:\; C^{\infty} \arrow \Lambda^1(M)\]
be the operator considered above. Then, there exists a unique induced%
\footnote{As usual, {\bf induced complex structure} means a complex
structure $L$ induced by a hyperk\"ahler structure, $L = aI + bJ + cK$,
$a,b,c \in \R$, $a^2 + b^2 + c^2 =1$.}
complex structure $L$, and an invertible element
$\alpha\in {{\Bbb H}\otimes_\R\C}$,
such that $\bar\6_L = \alpha(\bar \6 + tJ \circ \6)$.

\hfill

{\bf Proof:} Let 
\[ 
  {\cal D}= \text{Diff}^1\left(C^{\infty}(M), \Lambda^1(M)\right)
\]
be the space of all differential operators of first order
from $C^{\infty}(M)$ to $\Lambda^1(M)$.
The space ${\cal D}$ is equipped with a left action
of the quaternions, induced by the action of the 
quaternions  in $\Lambda^1(M)$.
Let $D\subset {\cal D}$ be the  subspace of ${\cal D}$
generated by $\bar \6_L$, $\6_L= \bar \6_{-L}$,
for all induced complex structures $L$. 
Clearly, $D$ contains the standard de Rham differential
$d = \frac{\6_L + \bar \6_L}{2}$. Let 
$d_L^c:= \frac{\6_L - \bar \6_L}{2\1}$ be the de Rham differential
twisted by $L$, which also lies in $D$. It is easy to check that
$d$, $d_I^c$, $d_J^c$, $d_K^c$ constitute a basis in $D$.
We identify $D$ with ${\Bbb H} \otimes_{\R} \C$, 
with $d$ going to $1\in \Bbb H$, and $d_\alpha^c$ going 
to $\alpha$, $\alpha=I, J, K\in \Bbb H$. Clearly, $D$ is preserved by 
the left action of the quaternions on ${\cal D}$, and 
is naturally isomorphic to the space
to ${\Bbb H} \otimes_{\R} \C= Mat_\C(2)$ as a representation of
$({{\Bbb H}\otimes_\R\C})^*= GL(2,\C)$. 
Let $C_0 \subset Mat_\C(2)$ be the set of all  $x\neq 0$
with $\det x =0$. Under the identification $D \stackrel i \arrow Mat_\C(2)$,
the operators $\6_L$ go to $C_0$, for all induced complex structures $L$.
Also, $i(\bar \6 + tJ \circ \6)$ belongs to $C_0$.
The quotient of 
$C_0$ by the left action of ${{\Bbb H}\otimes_\R\C}^*$
is $\C P^1$, and this $\C P^1$ is in a natural bijective correspondence
with the set of induced complex structures on $M$. 
Using \ref{_HxC_action_on_D_Remark_}, we obtain 
\ref{_6_new_induced_by_hype_expli_Claim_}. $\;\;\blacksquare$

\hfill

We have 
reduced \ref{_Tian_Todo_through_peri_Theorem_} to the following statement.

\hfill

\lemma \label{_Peri_of_bar_6+tJ6_Lemma_} 
Let $M$ be a hyperk\"ahler manifold, and $L$ be the induced 
complex structure constructed in 
\ref{_6_new_induced_by_hype_expli_Claim_}. Then the 2-form
$\Omega+ 2t \omega + t^2 \bar\Omega$ is of type (2,0) with
respect to $L$.

\hfill

{\bf Proof:} As with \ref{_6_new_induced_by_hype_expli_Claim_},
this statement is proven by an elementary linear-algebraic
computation. Let $W\subset \Gamma_M(\Lambda^2(M))$ be the $\C$-linear
space generated by $\omega_I$, $\omega_J$, $\omega_K$. 
Consider the standard action of $SU(2)$ on $W$. We obtain a 
representation

\[ 
    \rho:\; SU(2) \arrow End (W). 
\]
Consider the space $D$ of differential operators generated by $\bar \6_L$
for all induced complex structures (see 
\ref{_6_new_induced_by_hype_expli_Claim_}, Proof).
Then $D$ is canonically isomorphic to ${\Bbb H} \otimes_{\R} \C$.
The group ${{\Bbb H}\otimes_\R\C}^*=GL(2,\C)$ acts on 
$C_0\subset D={\Bbb H} \otimes_{\R} \C$ from the left and from the right,
with the left action being the same as we considered in 
\ref{_6_new_induced_by_hype_expli_Claim_} Proof.
We have identified the left quotient 
$(({{\Bbb H}\otimes_\R\C})^*) \backslash C_0$
with the set $\cal R$ of all induced complex structures. 
Let ${\Bbb H}^{un} = SU(2)$ be the group of unitary quaternions.
Clearly, the right action of 
${\Bbb H}^{un}\subset {\Bbb H}^* \subset {{\Bbb H}\otimes_\R\C}^*$
on $D$ induces the natural action of ${\Bbb H}^{un} = SU(2)$
on $\C P^1 = {{\Bbb H}\otimes_\R\C}^* \backslash C_0$.
In particular, under the right action of 
${\Bbb H}^{un}\subset {\Bbb H}^* \subset {{\Bbb H}\otimes_\R\C}^*$
on $C_0$, we have  
\begin{equation} \label{_right_ac_on_D_Equation_}
   g(\6_L) = \6_{g(L)}, \text{\ \ and\ \ } 
   g(\bar\6_L) = \bar\6_{g(L)} 
\end{equation}
where $L\in {\cal R}$ is the induced complex structure, and $g(L)$
is an image of $L$ under the natural action of 
${\Bbb H}^{un} = SU(2)$ on ${\cal R}= \C P^1$. 

\hfill

Let $P_c: \; {\cal R} \arrow {\Bbb P}(W)$ be the map associating
with an induced complex structure $L$ the line of all forms
$\lambda\in W$ which are of type $(2,0)$ with respect to $L$.
Clearly, for all $g\in SU(2)\otimes \C$, $L\in {\cal R}$, we have 
\begin{equation} \label{_SU(2)_commu_with_perio_Equation_}
   \rho(g)(P_c(L)) = P_c(g(L)),
\end{equation}
where $\rho:\; SU(2) \arrow End(W)$ is the representation defined above.
The space $W$ is equipped with a natural Hermitian metric
$(\cdot,\cdot)_{\c H}$, such that
$\omega_I$, $\omega_J$, $\omega_K$ is an orthonormal basis.
Whenever $\lambda\in W$ is a $(2,0)$-form for some induced
complex structure, we have $(\lambda,\lambda)_{\c H} =0$.
Let $C \subset {\Bbb P}(W)$ be the conic defined by 
$(\lambda,\lambda)_{\c H} =0$. This conic is naturally
isomorphic to $\C P^1$, so that the map $P_c:\; {\cal R} \arrow C$
is an isomorphism. 
Consider the Lie algebra $\g_0(\c H)\subset End(\Lambda^2(M))$ generated by 
$[L_{\omega_\alpha}, \Lambda_{\omega_\beta}]$, for $\alpha,\beta\in \{I, J,
K\}$, $\alpha\neq \beta$. From 
\eqref{_so5_relations_Equation_} it is clear that $\g_0(\c H)$
is naturally isomorphic to $\goth{su}(2)$.
Let $G_0(\c H)= SU(2)$ be the corresponding Lie group, acting
on the cohomology of $M$. Using the identification 
$G_0(\c H)\otimes \C = SU(2)\otimes\C = GL(2,\C)$, we may
consider $e^{t[L_{\omega_\alpha}, \Lambda_{\omega_\beta}]}$
as an element of $SU(2)\otimes\C = {{\Bbb H}\otimes_\R\C}^*$.
Let $p\in {{\Bbb H}\otimes_\R\C}^*$ be the element
which corresponds to $e^{t[L_\omega,\Lambda_{\bar\Omega}]}$.
By definition, \[ \rho(p) (\Omega) = \Omega+ 2t \omega + t^2 \bar\Omega.\]
On the other hand, a calculation shows that, 
under the natural right action of 
${{\Bbb H}\otimes_\R\C}^*$ on $D=  {\Bbb H} \otimes_{\R} \C$, 
we have $p(\bar\6) = \bar \6 + t J \6$. 
By \eqref{_right_ac_on_D_Equation_},
\eqref{_SU(2)_commu_with_perio_Equation_}, the vector

\[ 
   \rho(p)(\Omega)= \Omega+ 2t \omega + t^2 \bar\Omega
\]
belongs to the line $P_c(L)$. This implies that, for 
the induced complex structure $L$ of 
\ref{_6_new_induced_by_hype_expli_Claim_}, the differential
operators $p(\bar \6)$ and
$\bar\6_L$ define the same complex structure, where $p(\bar \6)$
means the image of $\bar\6\in D$ under the {\em right} action of 
${{\Bbb H}\otimes_\R\C}^*$. 
\ref{_Peri_of_bar_6+tJ6_Lemma_}, and consequently,
\ref{_Tian_Todo_through_peri_Theorem_}
 is proven. $\;\;\blacksquare$

\hfill


\section{Proof of Mirror Conjecture.}\label{_proof_mirro_Section_}


In this section, we prove \ref{_mirror_for_hype_Theorem_}. 
Let $\c B_1$ be an open ball in $H^{n-1,1}(M)$, $\phi:\; \c B_1 \arrow
Comp$ the Tian--Todorov map, $I = \phi(0)$. 
Let $\c B_2$ be an open ball in $H^{1,1}(M)$.
We identify $\c B_1$ with its image under $\phi$, considering $\c B_1$
as an open subset in $Comp$. 
We denote by $\c A_1$ the Yukawa VFA associated
with a trivialization of $K\restrict {\phi(\c B_1)}$
which we shall specify soon thereafter.
Let $\c A_2$ be a Quantum VFA, with base $\c B_2$. Shrinking
$\c B_1$, $\c B_2$ when necessary and using the natural 
isomorphism $H^{1,1}(M) \cong H^{n-1,1}(M)$, we obtain a canonical
linear isomorphism
\begin{equation} \label{_B_1_B_2_ide_Equation_}
   t:\; \; \c B_1 \arrow \c B_2. 
\end{equation}
To prove \ref{_mirror_for_hype_Theorem_}
we need to produce an isomorphism of VFA
$\theta:\; \c A_1 \arrow t^* \c A_2$,
where $t^* A_2$ is the pullback of $\c A_2$ under $t$.

\hfill

Let $C$ be, as usual, the quadric
hypersurface defined by $(x,x)_{\c H}=0$, and $P_c:\; Comp \arrow C$
be the period map.  Consider the
restriction $\calo(1)\restrict C$   
of $\calo(1)$ to $C \subset {\Bbb P}(H^2(M, \C))$.
Let $\Omega_s$ be the holomorphic line bundle on $Comp$,
$\Omega_s = P_c^*\left(\calo(1)\restrict C\right)$. 
Fix a holomorphic symplectic
form $\Omega$ on the complex manifold $(M, I)$.
We define a trivialization of the 1-dimensional complex space
 $\Omega_s \restrict {\phi(\c B_1)}$
as follows. Let $\alpha \in \c B_1$. Consider $\alpha$ as a $(1,1)$-form.
By definition, the line $P_c(\phi(\alpha))$
contains a vector $e^{[L_\alpha, \Lambda_\Omega]} \Omega\in H^2(M, \C)$,
where $e^{[L_\alpha, \Lambda_\Omega]}\in End(H^*(M))$ is the element
of $G^{I, 2}_0(M)$ corresponding to $\alpha\in H^{1,1}(M)$. 
Therefore, we may consider
$e^{[L_\alpha, \Lambda_\Omega]} \Omega$ as a vector of 
$\Omega_s \restrict{\phi(\alpha)}$. This gives a vector in the fiber
of $\Omega_s$ for every point $\phi(\alpha)\in \phi(\c B_1)$ and
trivializes $\Omega_s$ over $\phi(\c B_1) \subset Comp$.
Denote the section 
$\phi(\alpha)\arrow e^{[L_\alpha, \Lambda_\Omega]} 
\Omega\in \Omega_s \restrict{\phi(\alpha)}$ by $\nu$. 
Since $\Omega_s^{n/2} = K$, $\nu$ gives a trivialization of 
$K\restrict{\phi(\c B_1)}$. By $\c A_1$ we understand
Yukawa VFA associated with this trivialization.

\hfill

For the rest of this section, we
denote the endomorphism 
$e^{[L_\alpha, \Lambda_\Omega]}:\; H^*(M) \arrow H^*(M)$
by $e_\alpha$. The fibers of $\c A_1$, $\c A_2$ are naturally identified
with the total cohomology space $H^*(M)$. Let 
\[ 
   \theta_0:\; \c A_1 \arrow\c A_1
\] 
be the endomorphism acting
on the fiber $\c A_1\restrict{\phi(\alpha)}$ by $e_{-\alpha}$.
 Let 
\[ \eta_0:\; \c A_1 \arrow t^*\c A_2\] be the constant map
acting on the fibers of $\c A_1$, $\c A_2$ 
by \[ \eta_{I, \Omega}:\; H^*(M) \arrow H^*(M),\] where $\eta_{I, \Omega}$
is the Serre duality operator associated with $(I, \Omega)$
(Section \ref{_operator_Serre_dua_Section_}). 
Let \[ \theta= \theta_0\circ \eta_0:\; \c A_1 \arrow t^* \c A_2. \]
The following theorem proves the Mirror Conjecture for manifolds
of hyperk\"ahler type (\ref{_mirror_for_hype_Theorem_}).

\hfill

\theorem\label{_Mirror_proof_Theorem_}
The map $\theta$ induces an isomorphism of VFA.

\hfill

{\bf Proof:} By \ref{_eta_and_g(M)_Proposition_}, we have
$\eta_0[L_\alpha, \Lambda_\Omega]  \eta_0^{-1} = L_\alpha$.
Therefore, $\eta_0  e_\alpha \eta_0^{-1}$ is the map

\begin{equation} \label{e_alpha_conj_with_eta_Equation_}
\begin{split}
   \cdot \cup e^\alpha:\; H^*(M) &\arrow H^*(M), \\[2mm]
   \eta_0  e_\alpha \eta_0^{-1} (x) & = x\cup e^\alpha \\[2mm]
   x \cup e^\alpha =\sum \frac{x\cup \alpha^n}{n!}.
\end{split}
\end{equation}
Let $\nabla_m$ be the flat connection in $\c A_2$ associated
with the VFA structure, 
and $\gamma:\; \c B_2 \arrow H^*(M)$ be an arbitrary map
considered as a section of $\c A_2$.
Then $\gamma$ is parallel with respect to $\nabla_m$ if and only if
$\gamma(\alpha) = x\cup e^\alpha$, for some $x\in H^*(M)$, because
the operator $x \arrow x \cup e^{-\alpha}$ maps $\nabla_m$ to the 
constant connection $\nabla$ (see 
\ref{_multi_for_primary_Du_pote_Proposition_}). The operator
$\theta= \eta_0  \circ e_{-\alpha}$ maps the constant sections
$y$ of $\c A_1$ to the sections $\gamma(\alpha) = \eta_0(y)\cup e^\alpha$,
because $\eta_0  e_\alpha (y) = (\eta_0)^{-1}(y) \cup e^\alpha$ by
\eqref{e_alpha_conj_with_eta_Equation_}. In other words, $\theta$
maps constant sections of $\c A_1$ to the constant sections of 
$t^*\c A_2$. We obtain the following statement.

\hfill

\lemma \label{_theta_prese_conne_Lemma_} 
The map $\theta:\; \c A_1 \arrow t^* \c A_2$ commutes
with the flat connection on the variations 
of Frobenius algebras $\c A_1$, $t^* \c A_2$. 

$\;\;\blacksquare$

\hfill

To prove that $\theta$ is compatible with the Hodge filtration,
we make the following observation (\ref{_e_alpha_on_Hodge_filtra_Lemma_}).

\hfill

As in \eqref{_g_decompo_under_ad_I_Equation_}, consider the decomposition
\begin{equation}\label{_g_decompo_under_ad_I_(2)_Equation_}
   \g_0(M) \otimes \C = 
   \g_0^{I,-2}(M) \oplus\g_0^{I,0}(M) \oplus \g_0^{I,2}(M).
\end{equation}
Let $G_0^{I,-2}(M)$, $G_0^{I,0}(M)$, $G_0^{I,2}(M)\subset End(H^*(M))$
be the corresponding Lie groups. 

\hfill

\remark\label{_decompo_on_weights_G^I,i_o_Remark_} 
The multiplication 

\[ 
   G_0^{I,-2}(M)\times G_0^{I,0}(M) \times G_0^{I,2}(M) \arrow G_0(M)
\]
is surjective in a neighbourhood of unit in $G_0(M)$.

\hfill

\lemma \label{_e_alpha_on_Hodge_filtra_Lemma_} 
For a complex structure $L\in Comp$,
consider the Hodge filtration $F^0_L \subset F^1_L \subset ... $
on $H^p(M)$, 

\begin{equation}\label{_Hodge_fil_Equation_} 
  F^k_L := \bigoplus\limits_{i-q\leq k} H^{p-i,q}(M).
\end{equation}
associated with $L$. Then $e_{-\alpha}:\; H^p(M) \arrow H^p(M)$ maps 
$F^i_{\phi(\alpha)}$ to $F^i_I$.

\hfill

{\bf Proof:} Consider the standard action of a group $G_0(M)$ 
on \[ C \subset {\Bbb P} H^2(M, \C).\] By \ref{_Yu_equiv_Theorem_}, for
$I_1$, $I_2\in Comp$, $g\in G_0(M)$, such that 
\[ P_c(I_1)= g(P_c(I_2)), \]
we have \[ H^{p,q}_{I_1}(M) = g(H^{p,q}_{I_2}(M)).\] Since $G_0(M)$ acts
transitively on $C$, there exists
$g\in G_0(M)$ such that 
\[ g\left(H^{p,q}_I(M)\right)= H^{p,q}_{\phi(\alpha)}(M).\]
Shrinking $\c B_1$ if necessary, we may assume that $g$ admits
the decomposition considered in \ref{_decompo_on_weights_G^I,i_o_Remark_},
$g = g_{-2} g_0 g_2$, with $g_i \in G^{I, i}_0(M)$, $i = -2,0, 2$.
Under the natural action of $G_0(M) \otimes \C$ on 
$C \subset {\Bbb P}(H^2(M, \C))$, we have $g(P_c(I)) = P_c(\phi(\alpha))$.
Since $P_c(\phi(\alpha))$ is by definition a line which passes through
$e_\alpha(\Omega)$, we obtain that $g(\Omega)$ is proportional 
to $e_\alpha(\Omega)$. The group $G_0^{I,2}(M)$ stabilizes $\Omega$,
and $G_0^{I,0}(M)$ stabilizes the line passing through $\Omega$.
Therefore, $g_{-2}(\Omega)$ is proportional to $e_\alpha(\Omega)$.
The map $G^{I,-2}(M) \arrow C$, $g \arrow g(\Omega)$ is locally
an isomorphism (see Section \ref{_Periods_and_Tia_Todo_coordi_Section_}).
Therefore, $g_{-2} = e_\alpha$. 
By \ref{_eta_for_diff_c_H_equiv_Claim_}, the 
operator $g$ maps the Hodge grading
associated with the complex structure
$I$ to the Hodge grading associated with $\phi(\alpha)$.
To prove that the operator $e_\alpha$ is compatible with Hodge
filtration, we need to prove that
$g_0 g_2 = (g_{-2})^{-1} g$ preserves the Hodge filtration
associated with $I$. This is implied by the following general statement.

\hfill

\sublemma \label{_g_0_g_2_preserve_Hodge_fi_Sublemma_} 
Let $M$ be a compact holomorphically symplectic manifold of K\"ahler
type, $I$ its complex structure,
$G_0(M) \subset End(H^*(M))$ be the standard group acting
on its cohomology, $G_0^{I,-2}(M)$, $G_0^{I,0}(M)$, $G_0^{I,2}(M)$
be subgroups of $G_0(M)\otimes \C$ associated with $I$ and a decomposition
\eqref{_g_decompo_under_ad_I_(2)_Equation_}, and 
$s$ be any  element of $G_0^{I,0}(M)\times G_0^{I,2}(M)$.
Then the action of $s$ on $H^*(M)$ preserves the Hodge 
filtration \eqref{_Hodge_fil_Equation_} associated with
the complex structure $I$.

\hfill

{\bf Proof:} By definition, $G_0^{I,0}(M)$ preserves the Hodge
grading $H^*(M)= \oplus H^{p,q}_I(M)$. Therefore,
$ G_0^{I,0}(M)$ preserves the Hodge filtration. It remains
to show that $G_0^{I,2}(M)$ preserves the Hodge filtration.
The group $G_0^{I,2}(M)$ is by definition connected, 
and therefore we have to show that for all $\mu\in \g_0^{I,2}(M)$,
the action of $\mu$ on $H^*(M)$ preserves Hodge filtration.
By definition, $[\mu, ad\,I] = -2\1 ad\,I$. Therefore,
$ad\,I \left(\mu (\omega)\right) = (i - 2)\1 \mu (\omega)$, for all 
$\omega\in H^{p+i,p}(M)$. This means that 
$\mu (\omega) \in H^{p+i-1,p+1}(M)$. 
\ref{_g_0_g_2_preserve_Hodge_fi_Sublemma_} 
and consequently \ref{_e_alpha_on_Hodge_filtra_Lemma_}
is proven. $\;\;\blacksquare$

\hfill

\remark 
The group $G_0^{I,2}(M)$ manifestly does not preserve the
Hodge grading. Hence \eqref{_g_0_g_2_preserve_Hodge_fi_Sublemma_}
is false if we substitute ``filtration'' by ``grading''.

\hfill

\corollary \label{_theta_compa_Hodge_fi_Corollary_} 
The map $\theta:\; \c A_1 \arrow t^* \c A_2$ 
is compatible with Hodge filtration
associated with the VFAs $\c A_1$, $\c A_2$.

\hfill

{\bf Proof:} By definition of Serre duality, the
map $\eta_0$ maps the Hodge filtration
in $\c A_1\restrict{\phi(0)}$ to the Hodge filtration
in $\c A_2\restrict{0}$. Since the Hodge filtration
in $\c A_2$ is constant with respect to the trivial
connection $\nabla_0$, we can use
\ref{_e_alpha_on_Hodge_filtra_Lemma_}
and obtain \ref{_theta_compa_Hodge_fi_Corollary_}. $\;\;\blacksquare$

\hfill

Let $\c A_1^{gr}$, $\c A_2^{gr}$ be the associated graded bundles
for the variations of Frobenius algebras
$\c A_1$, $\c A_2$. The bundles $\c A_1^{gr}$, $\c A_2^{gr}$
are naturally equipped with a structure of weak VFA. By 
\ref{_theta_compa_Hodge_fi_Corollary_}, the map $\theta$ induces a map
of associated graded spaces
$\theta^{gr}:\; \c A_1^{gr}\arrow t^*\c A_2^{gr}$. To finish the proof
of \ref{_Mirror_proof_Theorem_}, we need to show only that
$\theta^{gr}$ is compatible with the weak VFA structure.

Consider the maps $\kappa_i:\; \c A_i^{gr}\otimes T \c B_i \arrow \c
A_i^{gr}$,  $\kappa(a\otimes t) = a\bullet t$, where 
$\bullet$ is the multiplication in VFA and $\tau_i$ is the
standard map which comes with the definition of VFA.
By definition, $\kappa_i$ is the Kodaira-Spencer map
associated with a 
weak $\C$-VHS. Since $\theta$ commutes with the weak $\C$-VHS structure
(\ref{_theta_compa_Hodge_fi_Corollary_}, 
\ref{_theta_prese_conne_Lemma_}), the following diagram is commutative:
\begin{equation} \label{_commu_dia:_theta_kappa_commu_Equation_}
\begin{CD} 
   T \c B_1 \otimes\c A_2^{gr} @>{\kappa_1}>> t^* \c A_2^{gr}\\
   @V{{dt} \otimes\theta^{gr}}VV  @VV {\theta^{gr}} V\\
   t^* T \c B_2\otimes\c A_1^{gr} @>{t^* \kappa_2}>> t^* \c A_1^{gr}
\end{CD}
\end{equation}
Assuming that $\theta^{gr}$ is an algebra homomorphism, we obtain
that the following diagram is also commutative:
\begin{equation} \label{_commu_dia:_theta_tau_commu_Equation_}
\begin{CD} 
   T \c B_1 @>{\tau_1}>> t^* \c A_2^{gr}\\
   @V{dt} VV  @VV {\theta^{gr}} V\\
   t^* T \c B_2 @>{t^* \tau_2}>> t^* \c A_1^{gr}
\end{CD}
\end{equation}
Therefore, to finish the proof of  \ref{_Mirror_proof_Theorem_},
we need only prove the following proposition.

\hfill

\proposition \label{_theta^gr_algebra_homo_Proposition_} 
Let $\theta^{gr}:\; \c A_1^{gr}\arrow t^*\c A_2^{gr}$ be the map constructed
above. Then $\theta^{gr}$ is compatible with the multiplicative structure
in $\c A_1^{gr}$, $\c A_2^{gr}$.

\hfill

{\bf Proof:} Consider the usual cohomology algebra 
$\left(H^*(M), \cup\right)$ of $M$. Then
$\c A_2^{gr}$ is a trivial bundle with a fiber 
$\left(H^*(M), \cup\right)$, and the multiplication
in $\c A_2^{gr}$ is compatible with this trivialization.
By definition of the Serre duality operator, the map
$\eta_0:\; \c A_1\restrict{\phi(0)} \arrow 
\left(H^*(M), \cup\right)$ commutes with the algebraic structure.
Therefore, to prove \ref{_theta^gr_algebra_homo_Proposition_},
we have to show that the map%
\footnote{By \ref{_e_alpha_on_Hodge_filtra_Lemma_}, $e_{-\alpha}$
 is compatible
with the Hodge filtrations on $\c A_1\restrict{\phi(\alpha)}$, 
$\c A_1\restrict{\phi(0)}$, associated with VFA.}
$e_{-\alpha}:\; \c A_1\restrict{\phi(\alpha)} \arrow \c
A_1\restrict{\phi(0)}$ induces a homomorphism of algebras

\[ 
  e_{-\alpha}^{gr}:\; 
  \left(\c A_1\restrict{\phi(\alpha)}\right)^{gr}
  \arrow \left(\c A_1\restrict{\phi(0)}\right)^{gr}.
\]

\hfill

Consider the natural action of the group $G_0(M)$ on $H^*(M)$. This action
induces an action of $G_0(M)$ on the tensor powers
\[ \left( H^*(M)\right)^{\otimes n} 
   \otimes\left( \left(H^*(M)\right)^*\right)^{\otimes n}. 
\]
Denote this action by $\lambda \arrow (\lambda)^g$. Let 
$I=\phi(0)$, $J= \phi(\alpha)$, and $\bullet_I$, 
$\bullet_J:\; H^*(M) \times H^*(M) \arrow H^*(M)$
be the Yukawa product maps in $\c A_1\restrict I$, $\c A_1\restrict J$.
Let $\cup:\; H^*(M) \times H^*(M) \arrow H^*(M)$ be the usual
cup-product. Consider $\bullet_I$, $\bullet_J$, $\cup$ 
as tensors over the space $H^*(M)$.
By \eqref{_Yukawa_via_eta_Equation_}, we have
\begin{equation} \label{_Yu_from_cup_Equation_}
 (\bullet_I) = (\cup)^{\eta_0}, \ \ (\bullet_J) = 
   (\cup)^{\eta_{\alpha}},
\end{equation}
where \[ \eta_0,\ \  \eta_\alpha \] are the Serre duality operators associated
with $\left(\phi(0), \Omega\right)$ and 
$\left(\phi(\alpha), e_\alpha(\Omega)\right)$. 
By definition, $\eta_\alpha = (\eta_0)^g$, where $g\in G_0(M)$
is the group element which defined in the proof of
\ref{_e_alpha_on_Hodge_filtra_Lemma_}. 
Therefore, \eqref{_Yu_from_cup_Equation_} implies that

\[ 
   (\bullet_J) = (\bullet_I)^g. 
\] 
Consider a decomposition $g = g_2 g_0 g_{-2}$
defined in the proof of \ref{_e_alpha_on_Hodge_filtra_Lemma_},
As we have seen, $e_{-\alpha}=  g^{-1} g_0 g_2$. 
The map 
\[ g^{-1}:\; \left( H^*(M), \bullet_J\right) \arrow 
   \left( H^*(M), \bullet_I\right)
\] 
induces an isomorphism
of Frobenius algebras and is compatible with the Hodge filtration.
Hence, $g$ induces an isomorphism 
on the associated graded algebras.
To prove that
$(g_2)^{-1}=e_{\alpha}:\; \left( H^*(M), \bullet_J\right) \arrow 
\left( H^*(M), \bullet_I\right)$ induces an isomorphism 
on the associated graded algebras, we need to show that
\[  g_0 g_2:\; \left( H^*(M), \bullet_I\right) \arrow 
    \left( H^*(M), \bullet_I\right)
\] 
induces an isomorphism 
on associated graded algebras. By
\ref{_stabi_I_commu_with_Yuka_Theorem_}, 
\[ g_0:\; \left( H^*(M), \bullet_I\right) \arrow 
   \left( H^*(M), \bullet_I\right)
\] is an algebra isomorphism.
Since $G_I^{I,0}(M)$ preserves the Hodge grading,
$g_0$ certainly induces an isomorphism 
on the associated graded algebras. It remains to show
that $g_2:\; \left( H^*(M), \bullet_I\right) \arrow 
\left( H^*(M), \bullet_I\right)$ induces an isomorphism 
on the associated graded algebras. This is implied by the
following little lemma, which finishes the proof of
\ref{_Mirror_proof_Theorem_}.

\hfill

\lemma\label{_G^I,2_acts_triv_on_asso_grad_Lemma_} 
Let $g_2\in G^{2,I}_0(M)$ be a group element and
\[ g_2:\; H^*(M) \arrow H^*(M)\] be the corresponding endomorphism
of $H^*(M)$. Consider the Hodge filtration on $H^*(M)$,
defined from the VFA structure on $\c A_1$ and
an isomorphism $H^*(M) \cong \c A_1\restrict{I}$.
\footnote{By definition of $\c A_1$, this
 filtration coincides with the standard Hodge
filtration on the cohomology space 
$H^*(M)$ associated with the complex structure $I$.}
Then $g_2$ acts as the identity on the associated graded
space.

\hfill

{\bf Proof:} The group $G_0^{2,I}(M)$ is by definition connected.
Thus, to prove that $g_2$ acts as the identity on the associated graded
space, we need to show that $\g_0^{2,I}(M)$ acts trivially on the 
associated graded space. In other words, for all
$\lambda \in \g_0^{2,I}(M)$, we need to show that
\begin{equation}\label{_lambda_decr_Hodge_fi_Equation_}
   \lambda \left(F^i_I H^*(M)\right) \subset F^{i-1}_I H^*(M), 
\end{equation}
where
$F_I^i$ is the Hodge filtration associated with $I$.
By definition of $\g_0^{2,I}(M)$, for all $\omega\in H^{p,q}_I(M)$,
we have $\lambda(\omega)\in  H^{p+1,q-1}_I(M)$. This proves 
\eqref{_lambda_decr_Hodge_fi_Equation_}. 
\ref{_G^I,2_acts_triv_on_asso_grad_Lemma_} is proven.
We have finished the proof of Mirror Symmetry for compact
holomorphically symplectic manifolds of K\"ahler type.
$\;\;\blacksquare$

\hfill

\hfill

{\bf Acknowledgements:} It is a pleasure to acknowledge fruitful discussions
with M. Bershadsky and A. Todorov, who explained to me the canonical 
coordinates on the moduli space. 
Also, B. A. Dubrovin was very kind to 
explain to me his work on quantum cohomology. 
I am grateful to F. Bogomolov,
D. Kazhdan, M. Kontsevich, A. Tyurin and S.-T. Yau 
for interest and encouragement. A. Beilinson and P. Deligne found
errors in the earliest versions of this work. Most of the thinking about
algebraic versions of Mirror Symmetry was done jointly with Valery  
Lunts. I am grateful to D. Kaledin, T. Pantev, L. Positselsky and
A. Vishik for most stimulating discussions and suggestions. 
Arthur Greenspoon made many invauable corrections to the manuscript.
Also, my gratitude is due to Julie Lynch and
International Press, who provided me with employment.

\end{document}